\documentclass[]{aastex631}

\usepackage{rotating}

\shorttitle{Kalliope \& Linus from ALMA and the VLA}
\shortauthors{de Kleer et al.}

\begin{document}

\title{Surface properties of the Kalliope-Linus system from ALMA and VLA data}

\author[0000-0002-9068-3428]{Katherine de Kleer}
\affil{California Institute of Technology \\
1200 E California Blvd M/C 150-21 \\
Pasadena, CA 91125 USA}

\author[0000-0001-6294-4523]{Saverio Cambioni}
\affil{Massachusetts Institute of Technology, Cambridge, MA 02139 USA}

\author[0000-0002-5344-820X]{Bryan Butler}
\affil{National Radio Astronomy Observatory, Albuquerque, NM 87106 USA}

\author[0000-0002-8441-2488]{Michael Shepard}
\affil{Commonwealth University of Pennsylvania, Bloomsburg, PA 17815, USA}

\begin{abstract}

The abundance and distribution of metal in asteroid surfaces can be constrained from thermal emission measurements at radio wavelengths, informing our understanding of planetesimal differentiation processes. We observed the M-type asteroid (22) Kalliope and its moon Linus in thermal emission at 1.3, 9, and 20 mm with the Atacama Large Millimeter/submillimeter Array (ALMA) and the Karl G. Jansky Very Large Array (VLA) over most of Kalliope’s rotation period. The 1.3 mm data provide $\sim$30 km resolution on the surface of Kalliope, while both the 1.3 and 9 mm data resolve Linus from Kalliope. We find a thermal inertia for Kalliope of 116$^{+326}_{-91}$ J m$^{-2}$ s$^{-0.5}$ K$^{-1}$ and emissivities of 0.65$\pm$0.02 at 1.3 mm, 0.56$\pm$0.03 at 9 mm, and 0.77$\pm$0.02 at 20 mm. Kalliope’s millimeter wavelength emission is suppressed compared to its centimeter wavelength emission, and is also depolarized. We measure emissivities for Linus of 0.73$\pm$0.04 and 0.85$\pm$0.17 at 1.3 and 9 mm respectively, indicating a less metal-rich surface composition for Linus. Spatial variability in Kalliope's emissivity reveals a region in the northern hemisphere with a high dielectric constant, suggestive of enhanced metal content. These results are together consistent with a scenario in which Linus formed from reaggregated ejecta from an impact onto a differentiated Kalliope, leaving Kalliope with a higher surface metal content than Linus, which is distributed heterogeneously across its surface. The low emissivity and lack of polarization suggest a reduced regolith composition where iron is in the form of metallic grains and constitutes $\sim$25\% of the surface composition. 

\end{abstract}


\section{Introduction} \label{sec:intro}

The Solar System's asteroid population consists of remnant planetesimals and their collisional fragments, enabling studies of the compositional diversity and chemical evolution of our Solar System's building blocks. While the majority of asteroids appear be similar to the primitive parent bodies of the carbonaceous and ordinary chondrite meteorites \citep{gaffey1978,gaffey1993}, a subset of the asteroids is thought to represent fragments from the interiors of planetesimals that melted and differentiated due primarily to heating from decay of $^{26}$Al and $^{60}$Fe \citep{urey1955,sahijpal2007}. Models and isotopic measurements of iron meteorites indicate that this differentiation occurred within a population of planetesimals that were at least $\sim$10 km in size within 3 Myr of formation of the calcium-aluminum inclusions \citep{merk2002,hevey2006,kruijer2017}. 

This scenario leads to predictions for the compositional variability expected among asteroids. For example, mantle material should be prevalent in the asteroid belt, and some asteroid families, or collections of fragments that share a parent body, should represent a range of compositions that reflects origination within different layers of a differentiated object. However, neither of these predictions is matched by the observations \citep{cellino2002,demeo2019}. This may indicate that most disrupted bodies were homogeneous, or perhaps that the fragments of a parent body are reaggregated mixtures of material originating from many depths within that parent body. In general, identifying and interpreting the remnants of planetesimal differentiation within the asteroid belt today has proven challenging.

It has been proposed that the remnant cores of the differentiated planetesimals make up the M-type asteroids or some subset of them \citep{cloutis1990}  (reclassified into the X complex in the \cite{bus1999} and \cite{demeo2009} taxonomies). This link is supported by their spectral resemblance to the iron meteorites; their relatively high bulk densities; and their often high radar albedos, which are attributed to high surface densities arising from high metal content. The target of our study --- (22) Kalliope (hereafter referred to as Kalliope) --- is the second largest M-type asteroid after (16) Psyche. Its high bulk density \citep[4.40$\pm$0.46 g cm$^{-3}$;][]{ferrais2022} indicates a metal-rich bulk composition. However, Kalliope's low radar albedo \cite[0.18$\pm$0.05, corresponding to a surface density of 2.3 g cm$^{-3}$;][]{shepard2015} indicates either a more porous or a more metal-poor surface than typical M-types. The difference between the bulk density and the surface density has been used to suggest that Kalliope may be a differentiated object whose crust and upper mantle have been largely stripped off, leaving it as a core dominated object like Mercury \citep{ferrais2022}. The presence of a large moon, Kalliope I Linus (hereafter referred to as Linus), and of the proposed existence of a collisional family, further indicate that Kalliope was disrupted via at least one major collision during its history \citep{merline2001,margot2001,broz2022}.

Kalliope's history of differentiation and major collisions make it one of the best candidates for exhibiting compositional variability across its surface and amongst its family members (potentially including Linus). \cite{broz2022} predict that the impact that formed Kalliope's proposed collisional family left Kalliope's surface unevenly covered with metallic ejecta, and their simulations indicate that the ejected material that should constitute members of the collisional family is mostly sourced from Kalliope's silicate mantle. This is consistent with Kalliope's high bulk density because a higher fraction of what is left behind is core material. However, the optical albedos and colors of the family members indicate that they are essentially all M types \citep{broz2022}, and \cite{laver2009} find that Linus is spectroscopically similar to Kalliope in the near-infrared and that neither target shows evidence for silicates. In addition, Kalliope is one of only two large asteroids out of nine for which \cite{marchis2012} did not find evidence for silicates in their emissivity spectra obtained from Spitzer, although subsequent near-infrared spectroscopy detected a spectral feature at 0.9 $\mu$m (but not 1.9 $\mu$m) \citep{ockert-bell2010,hardersen2011,neeley2014}. A hydrated silicate feature has also been tentatively identified on Kalliope around 2.7 $\mu$m \citep{rivkin2000,usui2019}.

Information on the abundance and distribution of metallic materials on planetary surfaces can also be obtained from the intensity and polarization of thermal emission observed at radio wavelengths. The emissivity, or amount of emission relative to a blackbody, is set primarily by the dielectric constant of the material on the surface and in the sub-surface, as is the fraction of thermal emission that is polarized (as a function of emission angle). The dielectric constant in turn is strongly dependent on the metal content of the material (in addition to the porosity). This is the same physical mechanism that allows metal content to be derived from the radar albedo of a surface, which is related to emissivity by Kirchhoff's law. At millimeter wavelengths, thermal emission can be directly spatially resolved with an effective linear resolution of $\sim$30 km in the main asteroid belt (${\sim}0.020\arcsec$ angular resolution) using the Atacama Large (sub-)Millimeter Array (ALMA) in its extended baseline configurations. At short centimeter wavelengths, which probe deeper layers in the sub-surface, objects in the main belt can be resolved with $\sim$100 km linear resolution (${\sim}0.065\arcsec$ angular resolution) with the Karl G. Jansky Very Large Array (VLA) in its most extended configuration. 

Millimeter and centimeter wavelength data together provide information on wavelength- and depth-dependent surface properties over the upper $\sim$1-10 cm of regolith, from which most of the thermal emission at these wavelengths originates. For binary asteroids with large secondaries and separations, the thermal emission from the primary and secondary can also be directly separated at both mm and cm wavelengths, and the thermophysical and dieletric surface properties of the two bodies can be compared to help determine the origin of the system. 
\cite{dekleer2021} and \cite{cambioni2022} used thermal emission and polarization observations from ALMA to constrain the metal content of the surface of (16) Psyche and detect spatial variations in surface properties. Here we apply the same observational and modeling approach to Kalliope and Linus using observations at 1.3 mm from ALMA. We also extend the analysis to include 9 and 20 mm observations from the VLA. At a spatial resolution of $\sim$30 km from ALMA and $\sim$100 km from the VLA, Kalliope is well resolved ($D_{eff}$=150 km) at 1.3 mm (5 resolution elements across) and partially resolved at 9 mm (1.5 resolution elements across). Linus is unresolved itself ($D$=28 km) but is separated from Kalliope sufficiently on the plane of the sky (of order ${\sim}0.5\arcsec$) to be easily distinguished in the observations. The comparison between the surface properties of Kalliope and Linus can provide new insight into whether Linus is composed of materials liberated from Kalliope by an impact.

The ALMA and VLA observations and data analysis are described in Sections \ref{sec:dataalma} and \ref{sec:datavla} respectively. The thermophysical and radiative transfer modeling is described in Section \ref{sec:models} and builds on \cite{dekleer2021} and \cite{cambioni2022} with the addition of several new robustness tests. The results are presented in Section \ref{sec:res} and discussed in Section \ref{sec:disc}. Conclusions are summarized in Section \ref{sec:conc}.

\section{Observations, data calibration, and imaging}\label{sec:obs}

\subsection{ALMA}\label{sec:dataalma}

Observations were made of Kalliope with ALMA, located in the Atacama Desert in Chile, on UTC 2019 June 20 between 03:12 and 06:28 UTC under project code 2018.1.01271.S. ALMA was in an extended configuration at the time of observation, with a maximum baseline of 15.2 km resulting in an angular resolution of ${\sim}20{\times}25$ milliarcseconds (mas). This corresponds to ${\sim}30{\times}40$ km on Kalliope, which was at a distance of 3.173 AU from the Sun and 2.165 AU from the Earth at the time of observation and subtended $\sim$100 mas. Full polarization data were collected in continuum mode, with four spectral windows each containing 64 channels over 2 GHz of bandwidth, centered at 224, 226, 240, and 242 GHz (around 1.3 mm, ALMA's Band 6). The observing set-up was identical to that of the (16) Psyche data published in \cite{dekleer2021}.

Kalliope was observed alternately with calibrators; the on-source time consisted of 93 scans each of duration 18-72 seconds, for a total on-source time of 75 minutes. Quasar J1718-2850 was used for calibration of complex gain as a function of time, J1751+0939 for polarization calibration, and J1924-2914 for bandpass and flux density scale calibration. Two checks were performed on the flux density scale calibration. The flux density of J1924-2914 as derived from the flux density scale calibrator was compared against direct measurements from the ALMA calibrator source catalogue, and was found to correspond well; the calibrator was quite stable in the two weeks preceding and following the observations, with variations below the 2\% level. Second, the flux density of J1751+0939 as calibrated by J1924-2914 was compared against direct measurements from the ALMA calibrator source catalogue, and was found to agree at better than the 2\% level. Given these checks, we adopt a flux density scale calibration uncertainty of 2\%. A larger uncertainty would translate into a larger uncertainty on our measured millimeter emissivity, but would not affect the comparisons between Kalliope and Linus because they are calibrated identically, and would not qualitatively affect the results. 

The raw data were calibrated via the ALMA pipeline. The data were delivered in the form of a calibrated Measurement Set (MS), which contains, for each target, the amplitude and phase of the cross-correlated signal between each antenna pair (referred to as ``visibilities''). Processing of the MS was performed using the Common Astronomy Software Applications (CASA) package \citep{mcmullin2007} as described in detail in \cite{dekleer2021} and briefly summarized below.

Because Kalliope rotates rapidly (${\sim}4.15$h period), the individual visibility scans capture it at different points in its rotation and capture Linus at different positions. Combining all visibility scans into one image would result in blended and blurred results unsuitable for further interpretation. We therefore break the data into snapshots, which are combinations of individual scans.  In principle, we could image each individual scan separately, but the signal-to-noise ratio (SNR) is low in that case; to increase the SNR we combine 3-5 individual scans into a snapshot.  The total duration of each snapshot never exceeded 5.5 minutes; this means that a location on the surface moves on the plane of the sky by less than half of the resolution size in all snapshots.  In that time Linus also only moves ${\sim}3$ masec in its orbit, a small fraction of the resolution.  An iterative imaging and self-calibration procedure \citep{cornwell1999} was performed on the visibilities in each snapshot to improve the phase calibration and produce an image for comparison with models. While self-calibration does not work well for a variable source, by choosing the snapshot durations as described above, we ensure that Kalliope is effectively not variable over a given snapshot. As with (16) Psyche, after the first round of self-calibration, further iterations with shorter solution intervals did not improve the SNR and were therefore not employed. Self-calibration solutions were derived from the total intensity data and applied to the polarization measurements, but no polarization signal was detected either before or after this procedure. The only practical difference between the procedure applied to these data and that described in \cite{dekleer2021} is that Linus was also imaged, by allowing the deconvolution algorithm to add model components in a circular region at Linus' location as seen in the images. The self-calibration procedure ultimately improved the SNR by $\sim$15\%, resulting in a SNR of $\sim$40-60 for each snapshot image and a typical image noise level of 1.5-2.0 K as estimated from the root-mean-square (rms) of the background in each image. Images were saved in Stokes $I$ (total intensity) and linear polarization intensity $P$, where $P=\sqrt{Q^2+U^2}$ for Stokes parameters $Q$ and $U$. The final images have a pixel scale of 3 milliarcseconds (mas) and an angular resolution of 20-25 mas. \par

The flux densities and relative positions of Kalliope and Linus were determined from the visibilities using the CASA package \textbf{uvmultifit} \citep{uvmultifit}, with Kalliope and Linus modeled as a disk and delta function respectively. The flux densities retrieved in this way were compared against those found by treating both components as disks in \textbf{uvmultifit}; by ``aperture photometry'' on the images (using \textbf{imstat} to sum up flux density in the images); by the OMFIT and TVSTAT tasks in the AIPS software package (OMFIT is similar to \textbf{uvmultifit}; TVSTAT is similar to \textbf{imstat}); and by summing the non-negative deconvolution model (CLEAN) components \citep{rau2011}. Summing the CLEAN components produced noisy light curves inconsistent with other the methods, while aperture photometry and the two-disk \textbf{uvmultifit} approach overestimate the flux density of Linus compared to the three other methods. The greatest consistency between methods, and the lowest scatter between observations, was found by using \textbf{uvmultifit} and modeling the brightness distribution of Linus as a delta function. This is also the preferred method for flux density retrieval if Linus is unresolved (discussed below), because it most accurately matches the true brightness distribution of the objects and fits the visibilities rather than the images, which are derived quantities. The reported uncertainties on the flux densities of the two objects, as well as their relative positions, are the uncertainties on the fit parameters as derived from the fitting algorithm. 

\begin{figure}[ht!]
\centering
\includegraphics[width=16cm]{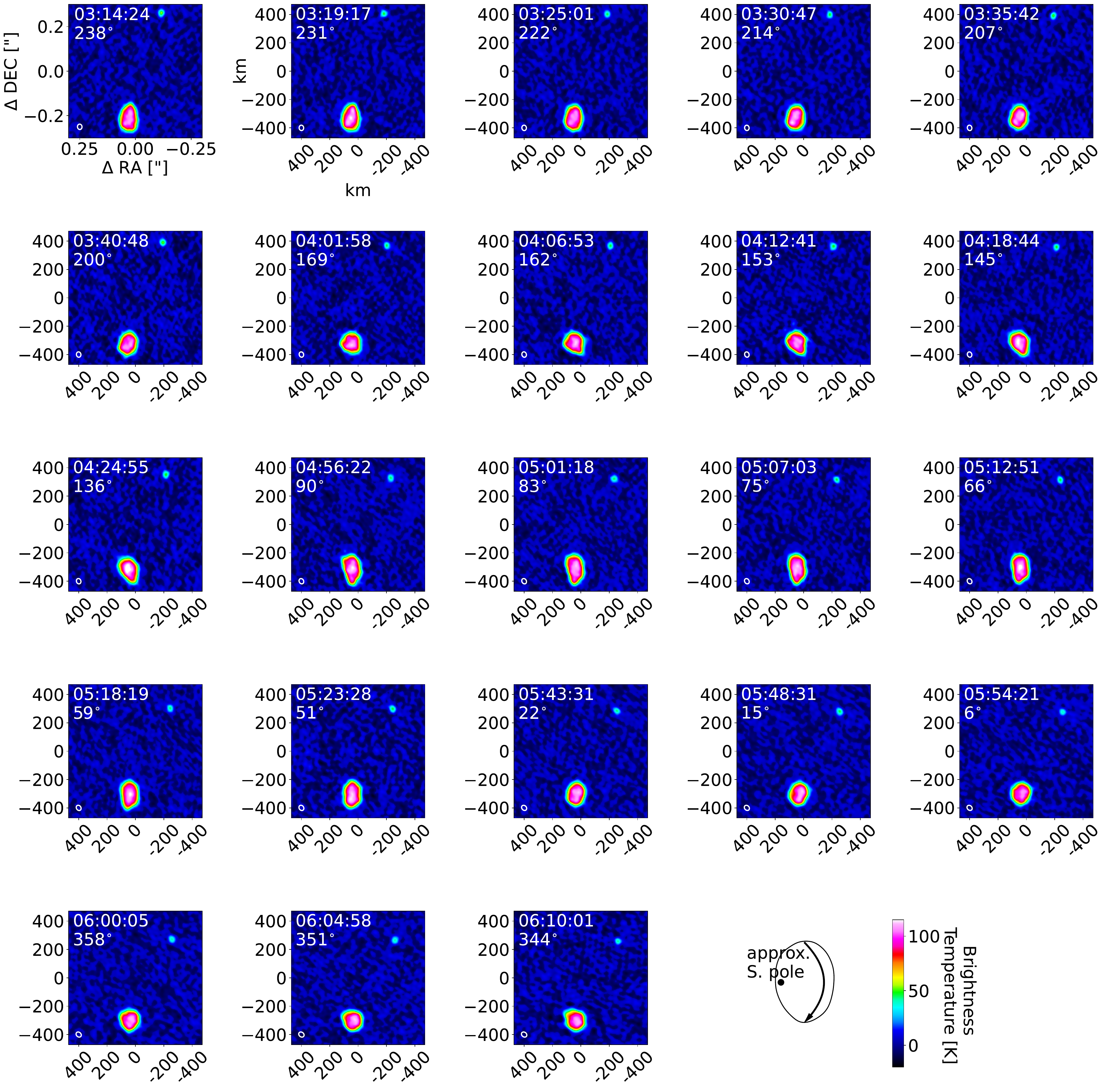}
\caption{Images of Kalliope and Linus over $\sim$3/4 of Kalliope's rotation period. Each image is labeled with the center time and the sub-observer longitude based on the SAGE shape model of \cite{ferrais2022}. The ellipse in the lower left corner indicates the size and shape of the resolution element. \label{fig:allims}}
\end{figure}

\begin{figure}[ht!]
\centering
\includegraphics[width=8cm]{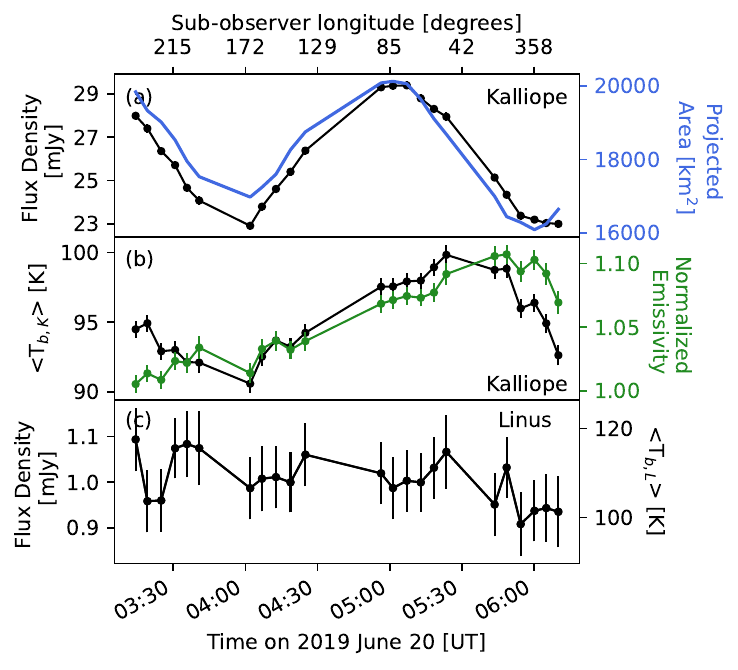}
\caption{Disk-integrated properties of Kalliope and Linus from ALMA. (a) Flux density of Kalliope plotted alongside its projected area according to the SAGE shape model from \cite{ferrais2022}. (b) The average brightness temperature of Kalliope $<T_{b,K}>$ according to these flux densities and areas; and (c) The flux density and average brightness temperature $<T_{b,L}>$ of Linus. Linus is $\sim$20 K warmer than Kalliope on average.\label{fig:diskint}}
\end{figure}

\begin{figure}[ht!]
\centering
\includegraphics[width=8cm]{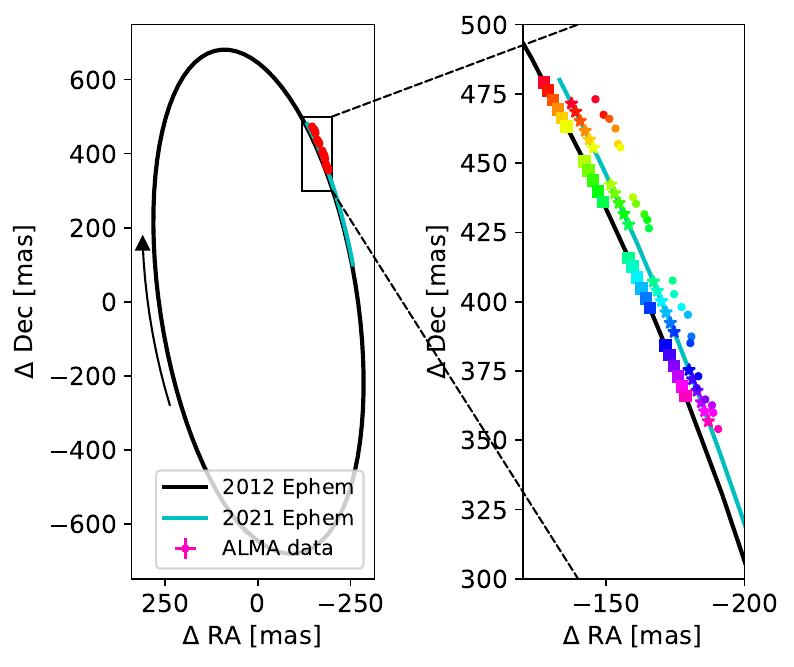}
\caption{Linus' orbit in RA and Dec offset in milliarcsecond from Kalliope from the ephemerides of \cite{vachier2012} and \cite{drummond2021} calculated at the time of the ALMA observations. In the zoomed panel, the squares and stars are the two ephemerides calculated at the exact time of the ALMA observation of the same color (light travel time corrected).} \label{fig:linusorbit}
\end{figure}

The flux densities and relative positions of Kalliope and Linus are presented in Table \ref{tbl:obs} for each snapshot, along with details of the observations. The uncertainties on the flux densities reported in Table \ref{tbl:obs} are the fit uncertainties only. In particular, the flux density scale uncertainty is not included in the uncertainties given in Table \ref{tbl:obs} because it affects all observations equivalently, but is incorporated whenever rotationally-averaged values are reported. The disk-averaged brightness temperature of Kalliope $<T_{b,K}>$ is calculated assuming the flux density arises from the emitting area predicted by the SAGE shape model of \cite{ferrais2022} for the midpoint time of each observation, and are corrected for the CMB emission blocked by the asteroid. The peak brightness temperature in each observation $T_{b,peak}$ and its uncertainty are computed from the peak flux density and rms of background image regions using the Planck function. The final set of calibrated images is shown in Figure \ref{fig:allims}, and Figure \ref{fig:diskint} plots the flux densities and disk-averaged brightness temperatures for both targets in each observation as a function of time and sub-observer longitude. The fitted position of Linus relative to Kalliope is plotted in Figure \ref{fig:linusorbit} along with orbits derived for Linus in prior work.\par

\begin{figure}[ht!]
\centering
\includegraphics[width=10cm]{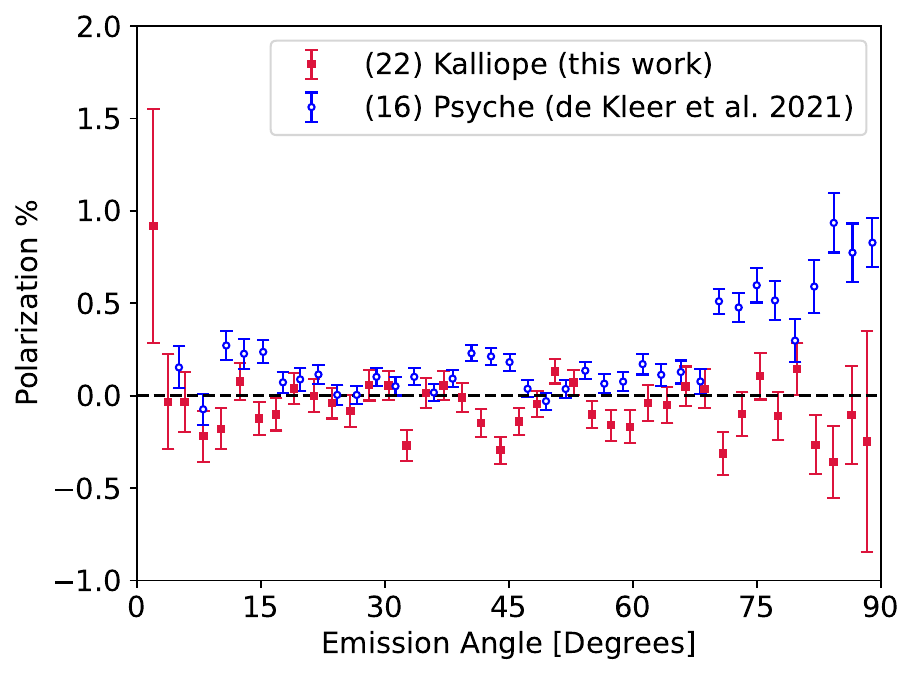}
\caption{Polarization fraction as a function of emission angle at 1.3 mm, binned across all ALMA observations. Kalliope shows no evidence for polarization. (16) Psyche is shown for comparison \citep[data from][]{dekleer2021}.}
\label{fig:polfrac} 
\end{figure}

No signal above the noise level is detected in any of the polarization images at all. Theoretically, the fraction of emission that is polarized is a function of emission angle. To provide the most sensitive test of polarization, we therefore mapped each pixel in each polarization image to its emission angle using the SAGE shape model of \cite{ferrais2022}, and binned the results into 40 emission angle bins; the result is shown in Figure \ref{fig:polfrac}. As in the images, no polarized emission is detected.

\begin{table}[h!]
\scriptsize
\begin{center}
\caption{ALMA observations on 2019-06-20 UT and derived properties of Kalliope and its moon Linus \label{tbl:obs}}
\begin{tabular}{ccccccccccccc}
\hline
\hline
\# & T$_{start}$ & T$_{end}$ & t$_{int}$ & Beam Size & CML$^a$ & $\Delta$RA & $\Delta$DEC & $T_{b,peak,K}$ & $<T_b>$ & F$_{tot,K}^b$ & F$_{tot,L}^b$ & F$_K$/F$_L$ \\
 & [UT] & [UT] & [sec] & [mas $\times$ mas] & [deg] & [mas] & [mas] & [K] & [K] & [mJy] & [mJy] & \\
\hline
1 & 03:12:06 & 03:16:42 & 200 & 20.8$\times$24.4 & 238 & -146.2$\pm$2.5 & 473.0$\pm$2.8 & 109.8 $\pm$ 1.6 & 94.4 $\pm$ 0.6 & 27.9 $\pm$ 0.1 & 1.09 $\pm$ 0.06 & 26 $\pm$ 2 \\ 
2 & 03:17:07 & 03:21:27 & 200 & 20.7$\times$24.2 & 231 & -149.1$\pm$2.9 & 467.4$\pm$3.3 & 114.7 $\pm$ 1.6 & 94.9 $\pm$ 0.6 & 27.4 $\pm$ 0.1 & 0.95 $\pm$ 0.06 & 29 $\pm$ 2 \\ 
3 & 03:22:25 & 03:27:37 & 200 & 20.5$\times$24.5 & 222 & -151.2$\pm$2.7 & 465.9$\pm$3.1 & 107.8 $\pm$ 1.5 & 92.9 $\pm$ 0.6 & 26.3 $\pm$ 0.1 & 0.96 $\pm$ 0.06 & 27 $\pm$ 2 \\ 
4 & 03:28:02 & 03:33:32 & 218 & 20.4$\times$24.2 & 214 & -153.4$\pm$2.3 & 462.4$\pm$2.6 & 113.5 $\pm$ 1.5 & 93.0 $\pm$ 0.6 & 25.7 $\pm$ 0.1 & 1.07 $\pm$ 0.06 & 24 $\pm$ 1 \\ 
5 & 03:33:34 & 03:37:51 & 181 & 20.4$\times$24.6 & 207 & -154.3$\pm$2.5 & 456.8$\pm$2.8 & 111.8 $\pm$ 1.6 & 92.1 $\pm$ 0.6 & 24.6 $\pm$ 0.1 & 1.08 $\pm$ 0.07 & 23 $\pm$ 2 \\ 
6 & 03:38:25 & 03:43:11 & 145 & 20.4$\times$24.4 & 200 & -155.2$\pm$2.9 & 455.6$\pm$3.3 & 109.8 $\pm$ 1.9 & 92.1 $\pm$ 0.7 & 24.0 $\pm$ 0.2 & 1.07 $\pm$ 0.08 & 22 $\pm$ 2 \\ 
7 & 03:59:36 & 04:04:21 & 200 & 20.1$\times$24.4 & 169 & -159.7$\pm$2.4 & 437.6$\pm$2.8 & 112.3 $\pm$ 1.4 & 90.5 $\pm$ 0.6 & 22.9 $\pm$ 0.1 & 0.98 $\pm$ 0.06 & 23 $\pm$ 2 \\ 
8 & 04:04:49 & 04:08:57 & 182 & 20.2$\times$24.9 & 162 & -160.9$\pm$2.6 & 435.3$\pm$2.9 & 113.0 $\pm$ 1.6 & 92.5 $\pm$ 0.6 & 23.7 $\pm$ 0.1 & 1.00 $\pm$ 0.07 & 24 $\pm$ 2 \\ 
9 & 04:09:58 & 04:15:25 & 200 & 20.0$\times$24.5 & 153 & -163.9$\pm$2.8 & 431.5$\pm$3.1 & 108.5 $\pm$ 1.6 & 93.6 $\pm$ 0.6 & 24.6 $\pm$ 0.1 & 1.01 $\pm$ 0.06 & 24 $\pm$ 2 \\ 
10 & 04:15:54 & 04:21:34 & 218 & 20.1$\times$24.6 & 145 & -164.9$\pm$2.4 & 429.5$\pm$2.7 & 116.0 $\pm$ 1.5 & 93.2 $\pm$ 0.6 & 25.4 $\pm$ 0.1 & 0.99 $\pm$ 0.06 & 25 $\pm$ 2 \\ 
11 & 04:22:02 & 04:27:48 & 200 & 20.2$\times$24.7 & 136 & -165.5$\pm$2.6 & 426.4$\pm$2.9 & 122.2 $\pm$ 1.6 & 94.2 $\pm$ 0.6 & 26.3 $\pm$ 0.1 & 1.06 $\pm$ 0.06 & 25 $\pm$ 2 \\ 
12 & 04:54:03 & 04:58:41 & 200 & 20.2$\times$24.9 & 90 & -174.0$\pm$2.7 & 407.6$\pm$2.9 & 114.6 $\pm$ 1.6 & 97.5 $\pm$ 0.5 & 29.3 $\pm$ 0.1 & 1.01 $\pm$ 0.06 & 29 $\pm$ 2 \\ 
13 & 04:59:07 & 05:03:29 & 200 & 20.2$\times$25.2 & 83 & -174.6$\pm$3.0 & 402.7$\pm$3.2 & 112.5 $\pm$ 1.4 & 97.5 $\pm$ 0.5 & 29.3 $\pm$ 0.1 & 0.98 $\pm$ 0.06 & 30 $\pm$ 2 \\ 
14 & 05:04:26 & 05:09:40 & 200 & 20.2$\times$25.1 & 75 & -177.2$\pm$2.5 & 398.0$\pm$2.8 & 112.7 $\pm$ 1.6 & 97.9 $\pm$ 0.5 & 29.4 $\pm$ 0.1 & 1.00 $\pm$ 0.06 & 29 $\pm$ 2 \\ 
15 & 05:10:06 & 05:15:36 & 218 & 20.2$\times$25.1 & 66 & -179.6$\pm$2.5 & 395.3$\pm$2.7 & 118.0 $\pm$ 1.4 & 97.9 $\pm$ 0.5 & 28.8 $\pm$ 0.1 & 0.99 $\pm$ 0.06 & 29 $\pm$ 2 \\ 
16 & 05:16:02 & 05:20:37 & 200 & 20.5$\times$25.4 & 59 & -180.8$\pm$2.6 & 387.4$\pm$2.8 & 115.7 $\pm$ 1.3 & 98.9 $\pm$ 0.6 & 28.3 $\pm$ 0.1 & 1.03 $\pm$ 0.06 & 27 $\pm$ 2 \\ 
17 & 05:21:35 & 05:25:22 & 145 & 20.7$\times$25.3 & 51 & -180.4$\pm$3.1 & 385.0$\pm$3.3 & 113.7 $\pm$ 1.8 & 99.8 $\pm$ 0.7 & 27.9 $\pm$ 0.2 & 1.06 $\pm$ 0.08 & 26 $\pm$ 2 \\ 
18 & 05:41:15 & 05:45:48 & 200 & 20.6$\times$25.8 & 22 & -183.3$\pm$3.0 & 373.1$\pm$3.0 & 111.7 $\pm$ 1.4 & 98.7 $\pm$ 0.6 & 25.1 $\pm$ 0.1 & 0.95 $\pm$ 0.06 & 26 $\pm$ 2 \\ 
19 & 05:46:13 & 05:50:50 & 218 & 20.7$\times$25.7 & 15 & -182.7$\pm$2.9 & 368.8$\pm$2.9 & 112.0 $\pm$ 1.4 & 98.8 $\pm$ 0.6 & 24.3 $\pm$ 0.1 & 1.03 $\pm$ 0.06 & 24 $\pm$ 2 \\ 
20 & 05:51:46 & 05:56:56 & 200 & 20.7$\times$25.9 & 6 & -185.8$\pm$3.3 & 364.7$\pm$3.3 & 107.7 $\pm$ 1.5 & 95.9 $\pm$ 0.6 & 23.3 $\pm$ 0.1 & 0.90 $\pm$ 0.06 & 26 $\pm$ 2 \\ 
21 & 05:57:21 & 06:02:49 & 218 & 20.7$\times$25.9 & 358 & -188.2$\pm$3.2 & 362.6$\pm$3.2 & 110.7 $\pm$ 1.3 & 96.4 $\pm$ 0.6 & 23.1 $\pm$ 0.1 & 0.93 $\pm$ 0.06 & 25 $\pm$ 2 \\ 
22 & 06:02:51 & 06:07:06 & 182 & 21.0$\times$26.3 & 351 & -188.7$\pm$3.4 & 359.9$\pm$3.4 & 111.9 $\pm$ 1.5 & 94.9 $\pm$ 0.7 & 23.0 $\pm$ 0.1 & 0.94 $\pm$ 0.07 & 24 $\pm$ 2 \\ 
23 & 06:07:31 & 06:12:31 & 163 & 21.1$\times$26.4 & 344 & -190.5$\pm$3.2 & 354.0$\pm$3.1 & 111.2 $\pm$ 1.5 & 92.6 $\pm$ 0.7 & 22.9 $\pm$ 0.1 & 0.93 $\pm$ 0.07 & 25 $\pm$ 2 \\ 
\end{tabular}
\end{center}
$^a$CML$=$Central meridian longitude, or sub-observer longitude, defined here relative to the best-fit ellipsoid. The subsolar longitude is 2$^{\circ}$ higher than the subobserver longitude, and the sub-observer and subsolar latitudes are -24$^{\circ}$ and -22$^{\circ}$ respectively in all observations.\\
$^b$Values in the table do not include a 2\% uncertainty on the flux density scale, which affect all observations identically. The $K$ and $L$ subscripts refer to Kalliope and Linus.
\end{table}

% KdK: Added 08/04/23 - updated 3/20/24

% KdK: Added 08/04/23

% KdK: Added 03/15/21 -- updated 8/12/24

% KdK: Added 11/02/23 ; Updated 5/12/24

\subsection{VLA} \label{sec:datavla}

Observations were made of Kalliope with the VLA, located on the Plains of San Agustin in New Mexico, USA, on two dates, under project code 20B-259.  Beginning on UTC 2020 December 18 at 23:07 and extending for 4.5 hours, observations at Ku-band (12-18 GHz) were executed, while Kalliope was at a distance of 2.677 AU from Earth and 2.662 AU from the Sun.  Beginning on UTC 2020 December 19 at 22:14 and also extending for 4.5 hours, observations at Ka-band (29-37 GHz) were executed, while Kalliope was at a distance of 2.701 AU from Earth and 2.662 AU from the Sun. The VLA was in its most extended configuration at the time (the ``A'' configuration), with maximum physical spacing between antennas of $\sim$36 km, resulting in a resolution of $\sim210\times125$ masec at Ku-band ($408\times243$ km linear resolution at the distance of Kalliope, so it was unresolved), and $\sim100\times60$ masec at Ka-band ($196\times118$ km linear resolution at the distance of  Kalliope, so it was slightly resolved).

Kalliope was observed alternately with calibrators in the normal way for VLA observations.  For the Ku-band observations, the on-source time consisted of 36 scans each of duration $\sim$320 seconds, for a total on-source time of $\sim$192 minutes.  For the Ka-band observations, the on-source time consisted of 60 scans each of duration $\sim$160 seconds, for a total on-source time of $\sim$160 minutes.  The calibrator J2331-1556 was used for calibration of complex gain as a function of time, and 3C 138 was used for delay, bandpass, polarization angle (data were taken in full polarization mode), and flux density scale calibration.  While 3C 138 is known to be variable \citep{perley2013,perley2017}, so not always appropriate for flux density scale calibration at the VLA, at the time it was thought it was stable and it was known that the other standard source near that right ascension (3C 48) was in the midst of a significant flare from OVRO monitoring \citep{richards2011}.  Unfortunately, it was later learned that 3C 138 was also undergoing a flare at the time, and our observations were in the early days of that flare\footnote{https://science.nrao.edu/facilities/vla/docs/manuals/oss/performance/fdscale}.  Therefore, we used the so-called ``Perley \& Butler 2017'' or ``PB17'' scale \citep{perley2017} for the flux density of 3C 138, but with an adjustment for that flare.  Internal VLA observations were used to derive the adjustments to the flux density of the source at Ku- and Ka-bands at the time of our observations, which were 7\% at Ku-band and 12\% at Ka-band (F. Schinzel \& B. Svoboda, personal communication).  Normally an uncertainty in the flux density scale of $\sim$1\% at Ku-band and $\sim$3\% at Ka-band would be adopted \citep{perley2013}; we adjust these up to 2\% at Ku-band and 5\% at Ka-band because of this additional uncertainty on the flux density of 3C 138.  As for the ALMA data, this systematic uncertainty does not affect the comparisons between Kalliope and Linus because they are calibrated identically, so does not affect those results. The flux density scale uncertainty is not included in the uncertainties given in Table \ref{tbl:VLAobs} because it affects all observations equivalently, but is incorporated when rotationally-averaged values are presented.

\begin{table}[h!]
\scriptsize
\begin{center}
\caption{VLA observations in Dec 2020 and derived properties of  Kalliope. \label{tbl:VLAobs}}
\begin{tabular}{cccccccccc}
\hline
Band & Freq & \# & T$_{mid}$ & Beam Size & CML$^a$ & $<T_b>$ & F$_{tot,K}^b$ & Pol \\
 & [GHz] & & [UT] & [mas $\times$ mas] & [deg] & [K] & [mJy] & [\%] \\
\hline
Ka & 33 GHz & 1 & 12-19 22:46:24 & 58$\times$110 & 146 & 113$\pm$4 & 0.503$\pm$0.019 & -3$\pm$8 \\
 & & 2 & 12-19 23:26:57 & 58$\times$97 & 87 & 122$\pm$3 & 0.550$\pm$0.014 & -5$\pm$5 \\
 & & 3 & 12-20 00:07:31 & 58$\times$90 & 29 & 114$\pm$3 & 0.497$\pm$0.013 & -2$\pm$4 \\
 & & 4 & 12-20 00:50:03 & 58$\times$88 & 327 & 111$\pm$3 & 0.496$\pm$0.013 & 3$\pm$4 \\
 & & 5 & 12-20 01:36:57 & 59$\times$89 & 259 & 113$\pm$3 & 0.518$\pm$0.013 & 6$\pm$4 \\
 & & 6 & 12-20 02:24:02 & 59$\times$97 & 191 & 106$\pm$3 & 0.471$\pm$0.012 & 3$\pm$5 \\
 \hline
Ku & 15 GHz & 1 & 12-18 23:38:58 & 124$\times$223 & 353 & 157$\pm$5 & 0.151$\pm$0.005 & \\
 & & 2 & 12-19 00:20:15 & 123$\times$208 & 293 & 157$\pm$5 & 0.155$\pm$0.005 & \\
 & & 3 & 12-19 01:01:32 & 123$\times$205 & 233 & 157$\pm$5 & 0.156$\pm$0.005 & \\
 & & 4 & 12-19 01:42:50 & 125$\times$208 & 173 & 148$\pm$5 & 0.144$\pm$0.005 & \\
 & & 5 & 12-19 02:24:07 & 127$\times$218 & 114 & 152$\pm$5 & 0.150$\pm$0.005 & \\
 & & 6 & 12-19 03:18:02 & 130$\times$250 & 36 & 141$\pm$5 & 0.134$\pm$0.005 & \\
 \end{tabular}
\end{center}
$^a$CML$=$Central meridian longitude, or sub-observer longitude, defined here relative to the best-fit ellipsoid. The subsolar longitude is 35$^{\circ}$ lower than the subobserver longitude, and the sub-observer and subsolar latitudes are 61$^{\circ}$ and 79$^{\circ}$ respectively in all observations.
$^b$Subscript $K$ refers to Kalliope. Values in the table do not include a 2\% (Ku-band) or 5\% (Ka-band) uncertainty on the overall flux density scale, which affect all observations identically.\\
\end{table}

For both bands, the raw data were calibrated with the VLA calibration pipeline \citep{kent2019} to produce a calibrated Measurement Set (MS), which contains the visibilities for each target.  At that point, CASA was used to export the visibilities to a UVFITS file which could be read into the AIPS (Astronomical Image Processing System) software package \citep{greisen2003}.  All further reduction and analysis were done in AIPS, described briefly below.

\begin{figure}[ht!]
\centering
\includegraphics[width=14cm]{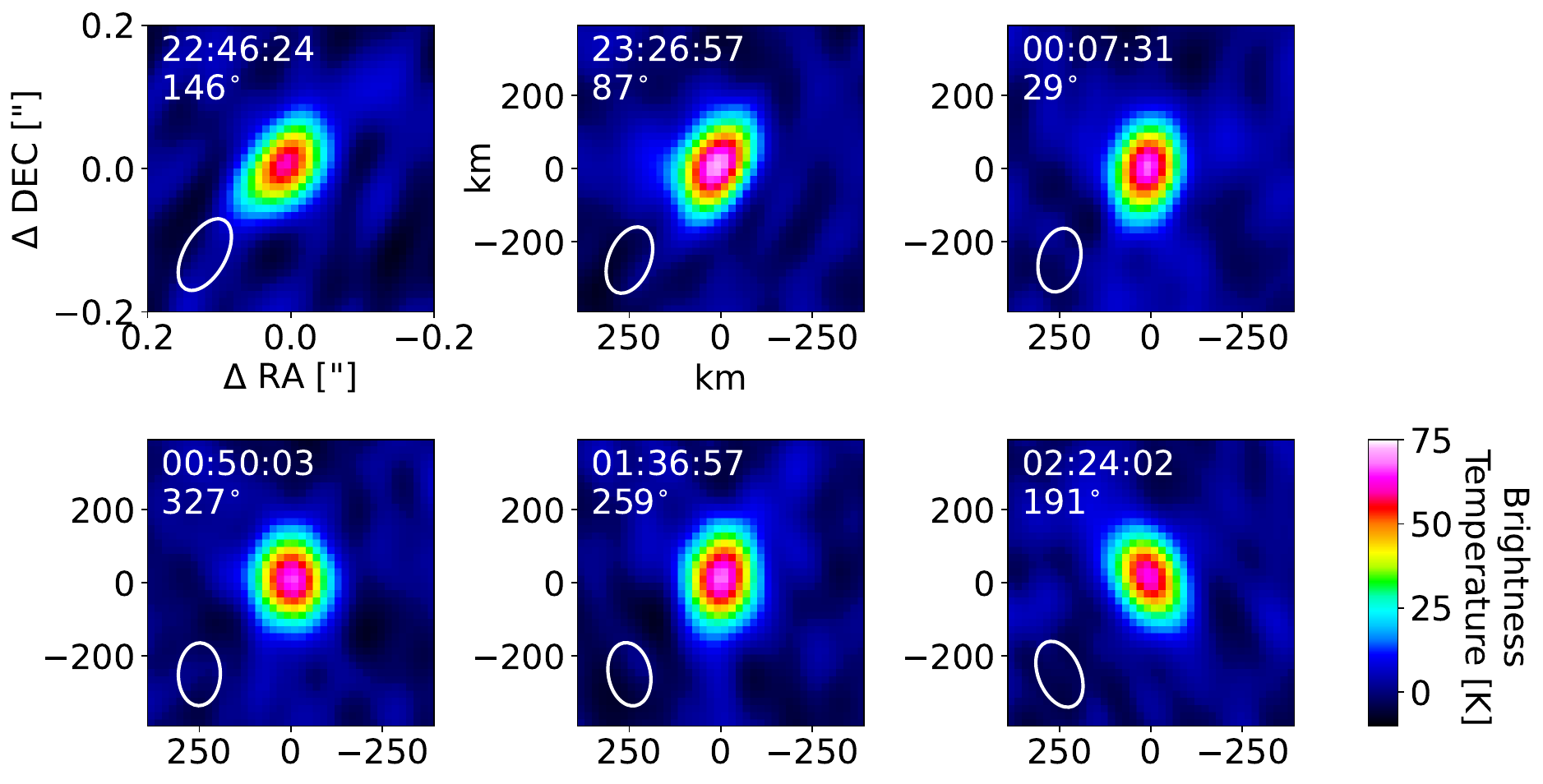}
\caption{VLA images of Kalliope at 9 mm (Ka band) on UT 2020 Dec 19 and 20. Markings and labels the same as in Figure \ref{fig:resims_global}.\label{fig:allKaims}}
\end{figure}

\begin{figure}[ht!]
\centering
\includegraphics[width=14cm]{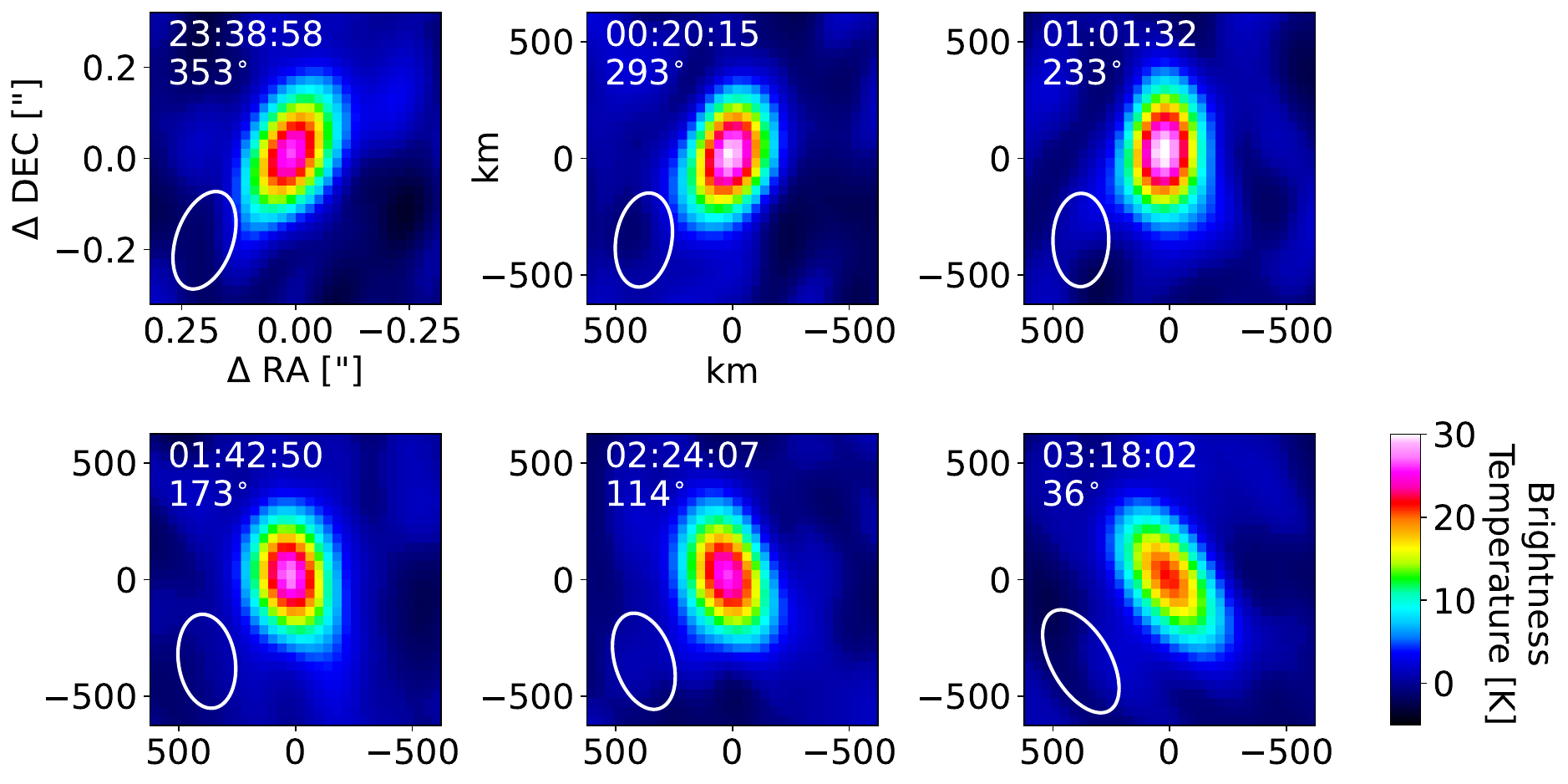}
\caption{VLA images of Kalliope at 2 cm (Ku band) on UT 2020 Dec 18 and 19. Markings and labels the same as in Figure \ref{fig:allims}.\label{fig:allKuims}}
\end{figure}

Similar to the ALMA data, and for the same reasons, individual scans were grouped together in snapshots with duration $\sim$45 minutes in both observations.  This interval was determined by constraining the motion of a point on the surface of Kalliope in a snapshot to be less than the smaller resolution dimension in the Ka-band data.  Although this has less meaning for the Ku-band observation because Kalliope is unresolved, it is a convenient time-scale for rotational differences to be manifested in the total flux density.  This resulted in six snapshots at both Ku- and Ka-bands.  Phase fluctuations during each scan on Kalliope result in a decorrelation of the data \citep{carilli1999}.  Fortunately the VLA continuously measures these fluctuations with a dedicated instrument, and those measurements can be used to correct for the decorrelation factor, which we did as the next step in the reduction.  We then imaged each snapshot, with a robust parameter of 1.0; these images are shown in Figure \ref{fig:allKaims} for Ka-band and Figure \ref{fig:allKuims} for Ku-band.  We attempted self-calibration (as for the ALMA data), but unfortunately there was not enough flux density from Kalliope at either Ka- or Ku-band to allow this to succeed.

% KdK: Added 08/20/23 - last updated 3/20/24

% KdK: Added 08/20/23 - updated 3/21/24

Flux densities were determined in a similar way and for the same fundamental reasons as for ALMA, i.e., we ended up using the fitted visibility zero-spacing flux density as our final flux density.  We did this using the OMFIT task in AIPS, with a slightly limb-darkened elliptical disk for Kalliope, as we did when using \textbf{uvmultifit} for the ALMA data (though \textbf{uvmultifit} does not allow for limb-darkening the difference is not significant in this case).  We did not include a point source for Linus in the fitted model, as the flux density from Linus in individual snapshots, even in Ka-band, is too faint to be detected.  The same cross-comparisons were made as were done for the ALMA analysis (TVSTAT, IMSTAT, counting flux density in CLEAN components) with similar results. The fitted flux densities and derived brightness temperatures are shown in Table \ref{tbl:VLAobs}.  The reported uncertainties on the flux densities for Kalliope are the formal uncertainties as derived from the fit.

As in the ALMA case, the disk-averaged brightness temperature of Kalliope $<T_{b,K}>$ is calculated assuming the flux density arises from the emitting area predicted by the SAGE shape model of \cite{ferrais2022} for the midpoint time of each observation. The peak brightness temperature in each observation $T_{b,peak}$ and its uncertainty are computed from the peak flux density and rms of background image regions using the Planck function.

At Ku-band the flux density from Linus is too faint to be detected in the time we observed.  As noted above, even at Ka-band, the flux density is too faint to be detected in the individual snapshots.  However, by shifting and co-adding all of the data together it can actually be detected.  We did this by first subtracting the emission from Kalliope from the visibilities in each snapshot (since the viewing geometry is changing for each snapshot), then adding all of the data back together, then adjusting the phase center of the visibilities to track Linus instead of Kalliope throughout the observation \citep{butler1999} and imaging that resulting dataset.  Linus can clearly be seen in the resulting image, and the resulting flux density is given in Table \ref{tbl:VLAobs}.

The VLA calibration pipeline does not perform polarization calibration.  We performed this calibration ourselves for the Ka-band data (for the unresolved Ku-band data we expect no detectable polarized emission), in the standard way in AIPS, using J2331-1556 as the so-called ``D-term'' calibrator and 3C 138 as the absolute polarization angle calibrator (its flare did not affect its polarization angle in a significant way).  As for the ALMA data, polarization intensity images were formed for each snapshot.  Also as for the ALMA data, we see no polarized emission, to a one-sigma level of ${\sim}5\%$ in any snapshot (individual snapshot limits are shown in Table \ref{tbl:VLAobs}), or a one-sigma level of ${\sim}2\%$ for the combined snapshots.

\section{Models\label{sec:models}}

We first compute upper and lower emissivity bounds for the surface of Kalliope by analyzing its thermal emission under the assumptions of zero and infinite thermal inertia, $\Gamma$ (SI units of J m$^{-2}$ s$^{-0.5}$ K$^{-1}$ are assumed throughout, sometimes represented with `tiu' for `thermal inertia unit'), defined as the resistance of the surface material to change temperature with changing insolation (Section \ref{sec:emissivity_bounds}). Next, we refine these bounds by means of a thermophysical model. For the 1.3 mm data, which is observed at high spatial resolution, we generate thermal emission images at the same viewing geometry and wavelength $\lambda$ of the observations with $\Gamma$ and the dielectric constant of the surface material, $\epsilon$, as free parameters (Section \ref{sec:TPM}). We perform fits using two versions of the thermophysical model: a ``global model'' in which we assume that Kalliope's surface has a uniform value of $\Gamma$ and $\epsilon$ across the surface (Section \ref{sec:global_TPM}; \cite{dekleer2021}); and a ``local model'' in which we allow $\Gamma$ and $\epsilon$ to vary across the surface (Section \ref{sec:local_TPM}; \cite{cambioni2022}). The ``local model'' takes advantage of the fact that each region of the surface is observed at several times of day, which breaks the degeneracy between $\Gamma$ and $\epsilon$ that exists when a region is only observed at one time of day. We also present new sensitivity tests in which we explore the sensitivity of the results derived from the global model to the signal-to-noise of the data, the treatment of emissivity, the shape model, and the electrical properties of the surface (Section \ref{subsec:TPM_tests}). For the 9--20 mm data, the thermophysical model is run for the observing geometry of the observations, adopting the best-fit thermal inertia from the ALMA data. The emissivity of the 9--20 mm observations is calculated as the ratio of the observed disk-averaged brightness temperature to the average model surface temperature on the observed hemisphere, where the observed disk-averaged brightness temperature is calculated from the disk-integrated flux density combined with the projected area of the asteroid at the time of observation from the shape model.

\subsection{Emissivity bounds}
\label{sec:emissivity_bounds}

To give some simple intuition prior to the more complex modeling, we calculate the bounds on the expected surface temperature of Kalliope by approximating the full plausible range of subsolar temperatures using the standard thermal model \citep[STM,][]{lebofsky1986}  and fast-rotating model \cite[FRM,][]{lebofsky1989}. The STM and FRM represent the nonrotating (or zero thermal inertia) and fast-rotating (or infinite thermal inertia) end members, respectively, and can be considered the absolute upper and lower limits on the true peak surface temperature of Kalliope. Our estimates assume a geometric albedo of 0.166$\pm$0.005 \citep{2019PDSSMainzer}, a phase integral of 0.42$\pm$0.02 appropriate for moderate albedo asteroids \citep{shevchenko2019}, and a bolometric emissivity of 0.8--0.9 characteristic of metal or silicate surfaces.

The disk-averaged brightness temperatures for the VLA observations (9-20 mm) are in the range of 105--155 K (Table \ref{tbl:VLAobs}), higher than in the 1.3-mm data (90--100 K; Table \ref{tbl:obs}). At the time the VLA data were taken (UT 2020 Dec 18-20) Kalliope was closer to the Sun with a Sun-target distance of 2.724 AU (compared to 3.173 AU for the 1.3 mm data). In addition, when the VLA data were taken Kalliope's north pole was facing the Sun (subsolar latitude of 79$^{\circ}$) while during the ALMA observations its equator faced the Sun (subsolar latitude of -22$^{\circ}$). Thus the actual surface temperature on the sunlit side of the asteroid was higher at the time of the VLA observations, consistent with the higher brightness temperatures measured with the VLA. In the VLA observing geometry, the sunlit side of the asteroid therefore receives insolation over the whole asteroid rotation with no opportunity to re-radiate at night, and hence the STM case is a good approximation of the physical surface temperature. Using the STM with beaming parameters of $\eta=$0.75--1.0 to encompass differing prior definitions \citep{lebofsky1986,lebofsky1989}, the expected flux densities from Kalliope's surface are 860--920 $\mu$Jy at 9 mm and 180--195 $\mu$Jy at 20 mm. A comparison to the VLA measurements of 501$\pm$29 and 148$\pm$6 $\mu$Jy for the two wavelengths respectively gives emissivity estimates of 0.55-0.59 at 0.9 cm and 0.76-0.81 at 2.0 cm.

The 1.3-mm observations have sufficient spatial resolution that the peak measured surface temperature can be compared with those predicted by the STM and FRM. The modeled subsolar temperature at the time of the ALMA observations is 222-247 K for the STM and 167-173 K for the FRM. Based on a peak brightness temperature of 122 K observed with ALMA, the 1.3-mm emissivity is 0.50--0.55 using the STM and 0.71--0.73 using the FRM. To derive the thermal inertia of the surface and hence a more precise constraint on emissivity, we employ the thermophysical model, as described in the next section.

\subsection{Thermophysical model}
\label{sec:TPM}

We generate thermal emission images at the viewing geometry and wavelength $\lambda$ of the observations. The free parameters in the model are the thermal inertia $\Gamma$, and the real part of the dielectric constant $\epsilon'$ for complex dielectric constant $\tilde{\epsilon} = \epsilon' + i\epsilon''$. As in \citet{dekleer2021} and \citet{cambioni2022}, we make the simplification that $|\tilde{\epsilon}| = \epsilon'$ (which we refer to simply as $\epsilon$ hereafter) because $\epsilon''$ is a negligible contribution to $|\tilde{\epsilon}|$ for most meteorites \citep{campbell1969}. We summarize the model below, specifying the model assumptions and a few updates with respect to the previous versions that are described in detail in \citet{dekleer2021} and \citet{cambioni2022}. The results are presented in Section \ref{subsec:TPM_results}.

\subsubsection{Global model}
\label{sec:global_TPM}

For a given epoch of observation, we solve the 1-dimensional diffusion equation with thermal inertia $\Gamma$ as free parameter using the well-established ThermoPhysical Model (TPM) from \cite{delbo2015}. We assume large-scale topography as described by the SAGE shape model from \citet{ferrais2022} with the corresponding spin pole orientation (in DAMIT\footnote{https://astro.troja.mff.cuni.cz/projects/damit/pages/documentation} notation: ecliptic latitude 196.7$^\circ$, ecliptic longitude 2.7$^\circ$, rotation period $P$ = 4.14819911 hours, and initial rotation angle $\phi_0$ = 334.0 at epoch $t_0$ = 2446324.768 julian days) with no surface roughness \citep{broz2023}. In Section \ref{subsec:TPM_results}, we explore the robustness of the results to the choice of the shape model and show that the SAGE shape model yields a better goodness of fit than the MPCD-ADAM shape model from \citet{ferrais2022} (earlier shape models are visually a poor match and were not tested or used). The \citet{ferrais2022} shape models are uniquely well suited to our observations, since they are based in part on high-resolution adaptive optics data obtained during the weeks bracketing the ALMA observations. We assume a geometric albedo of 0.166 \citep{2019PDSSMainzer} corresponding to a Bond albedo of $\sim$0.07, and a bolometric emissivity equal to 0.9. 

The TPM yields temperatures as a function of depth for each facet, at the time and viewing geometry of each observation, in the same way they would be calculated for interpretation of infrared data. However, whereas a thermal model used to interpret infrared data would compute the thermal emission from each facet based on its surface temperature then sum the facet emission together, for resolved millimeter data we need to account for subsurface emission and also maintain the spatial information on the emission coming from each facet. We therefore use the output of the TPM to compute the flux density $I_\nu$ for each facet of the shape model by integrating the emission $I_\nu(z)$ as a function of depth, $z$ (calculated from $T(z)$ for that facet) within the subsurface:

\begin{equation}
I_\nu = \frac{\int_0^\infty~I_\nu(z)~e^{-z/(\delta_{elec}\cos{\theta_t})}dz}{\int_0^\infty~e^{-z/(\delta_{elec}\cos{\theta_t})}dz}
\end{equation}

\noindent
where $\delta_{elec}$ is the electrical skin depth (the depth over which the emission is attenuated by $1/e$), and $\cos{\theta_t} = (1-\frac{sin^2{\theta}}{\epsilon'})^{1/2}$, where $\theta$ is the emission angle between the direction normal to the facet and the direction to the observer. The electrical skin depth depends on the imaginary part of the refractive index, $\kappa$, following $\delta_{elec} =\lambda/(4\pi \kappa)$. For modeling of (16) Psyche, \cite{dekleer2021} and \cite{cambioni2022} adopted $\delta_{elec}=$ 2 mm (at a wavelength of 1.3 mm) under the assumption of a metal-rich surface. Kalliope's lower radar albedo suggests a surface that is less metal-rich than Psyche's, and we therefore adopt a value of $\delta_{elec}=$10$\lambda$. This corresponds to $\kappa \sim$0.01, which is roughly in the middle of the range for the powdered rock and meteorite samples measured by \cite{campbell1969}. We explore the effect on our results of the assumed $\delta_{elec}$ in Section \ref{sec:tests_delec}. In the model, we parameterize the depth $z$ in units of diurnal thermal skin depth $\delta_{th}$, the depth over which the amplitude of the diurnal temperature wave is reduced by $1/e$:

\begin{equation}
\delta_{th} = \sqrt{\frac{P}{\pi}} \frac{\Gamma}{\rho c_p}
\end{equation}

\noindent where the surface bulk density $\rho$ is derived from the radar albedo using Eq. 5 in \citet{shepard2015}, based on the radar measurements from \citet{2007IcarMagri}. The heat capacity $c_p$ is assumed to be equal to 500 J/kg K, which is an average value for meteorites \citep{2021JGREPiqueux}. The thermal emission from each facet is multiplied by the millimeter emissivity E($\theta$) of the surface material for its emission angle $\theta$, computed from the Fresnel reflection coefficients for polarization parallel and perpendicular to the direction of propagation, which depends on dielectric constant of the material, a free parameter in the model \citep{dekleer2021}. 

Next, we obtain model thermal-emission images for Kalliope by projecting the emission from each facet onto the plane-of-sky in the International Celestial Reference System (ICRS) using the procedure described in \citet{cambioni2022}. This procedure consists of four steps: (1) rotation of the facets of the shape model from asteroid-centric coordinates to ICRS; (2) projection of the rotated facets onto the plane-of-sky, taking into account the projected area of the surface facets; (3) interpolation of the thermal emission values of the projected facets that are visible to the telescope in RA and Dec; and 
(4) convolution of the image with the telescope beam. Finally, we compute the goodness of fit of the model images $M$ with respect to the data images $D$ across all epochs as a function of $\Gamma$ and $\epsilon$ in units of $\chi_r^2$:

\begin{equation}
\label{eq:chi_squared}
    \chi_r^2= \frac{1}{N_{pxl}-N_{par}} \sum_{i=1}^{N_{obs}} \sum_{j=1}^{N_{pxl}^i}\bigg(\frac{M_j^i-D_j^i}{\sigma^i}\bigg)^2
\end{equation}
\noindent
where $N_{pxl}$ is the total number of pixels across Kalliope, $N_{obs}$ is the number of images, and $N_{par}$ is the number of free parameters in the model. The noise level $\sigma^i$ associated with the \textit{i}th image is the image rms described in Section \ref{sec:dataalma}. As in previous studies \citep{cambioni2022}, we consider that every combination of ($\Gamma$, $\epsilon$) that satisfies 

\begin{equation}
\label{eq:chi_2_criterion}
    \chi_r^2 < \chi_{r,min}^2 \bigg[1+\sqrt{2/(N_{dat}-N_{par})}\bigg]
\end{equation}
\noindent
is an admissible solution, where $N_{dat}$ is the number of total pixels on Kalliope divided by the number of pixels per resolution element. This definition of $N_{dat}$ accounts for the fact that the actual resolution of the data is oversampled by the number of pixels per resolution element. The latter is equal to 65 for the ALMA dataset. In the results, we present the minimum-$\chi^2$ model as the best-fit value. The reported 1$\sigma$ uncertainties are the 68th percentile above and below the minimum-$\chi^2$ of the full range that satisfies the criterion of Equation \ref{eq:chi_2_criterion}.

To fit the $N_{obs}$ = 23 ALMA thermal images, we generated forward models with $\Gamma=$ [5, 10, 15, 20, 25, 50, 75, 100, 116, 135, 156, 181, 210, 244, 283, 328, 381, 442, 512, 594, 690, 800], $\epsilon_1$ between 0 and 60 with a step size of 1 and $\epsilon_2$ of [2, 5, 10, 15, 20, 25, 30, 35, 40, 45, 50, 55, 60] assuming a smooth surface. The motivation for and application of the two dielectric constant parameters $\epsilon_1$ and $\epsilon_2$ in the model are described in Section \ref{sec:emiss_test}.

\subsubsection{Sensitivity tests on the global model}
\label{subsec:TPM_tests}

We run a series of robustness test aimed at exploring the sensitivity of the solution found by the global model fits to the quality of the data and to model assumptions. We use the results of these tests to guide our  global model fitting. 

\paragraph{SNR of the data}\label{sec:test_SNR}
Our first test explores the effect of the SNR of the data on the accuracy of the retrieved parameters. To investigate this, we generated a dataset of fake ALMA data images by taking model images for several choices of ($\Gamma$, $\epsilon$) and adding noise with realistic characteristics at different levels. In practice, the noise that was used was simply the Stokes Q and U images (since they contained no signal) scaled up or down by different amounts and added to the model images. The fake dataset was produced by one of the authors, and sent to another author who performed the retrieval using the thermophysical model without knowledge of which ($\Gamma$, $\epsilon$) choices had been made in generating the fake data. We attempted the retrieval of $\Gamma$ and $\epsilon$ from nine different sets of fake data with different ($\Gamma$, $\epsilon$) choices and SNR ranging from $\sim$ 10 to $\sim$ 50. In all cases, the retrieved $\Gamma$ and $\epsilon$ were within 1$\sigma$ of the true values. We therefore find that our global TPM fits are robust to the SNR of the data within a range of SNR values ($\sim$10 to $\sim$50).

\paragraph{Emissivity parameterization}\label{sec:emiss_test}
We test the effect of how we model emissivity in the thermophysical model. In the standard version of our model described in Section \ref{sec:global_TPM}, the dielectric constant $\epsilon$ controls both the normal emissivity of the surface $E(0)$ as well as how the emissivity varies with emission angle $E({\theta})/E({0})$. However, for asteroid (16) Psyche, \citet{dekleer2021} showed that this assumption led to systematic high residual temperatures at the limb of the asteroid as imaged by ALMA, and the parameters controlling these two effects needed to be decoupled to match the data. This decoupling in effect parameterizes the role of surface roughness and particle scattering in controlling how the emission drops off with emission angle.

We explore what level of error is introduced by assuming that one value of dielectric constant controls both $E(0)$ and $E({\theta})/E({0})$ by producing model images where these parameters are decoupled, and retrieving them using a model where they are coupled. To do this, we again created a dataset of fake ALMA data images in which the value of dielectric constant $\epsilon_1$ that controls $E({0})$ may be different from the value $\epsilon_2$ that controls $E({\theta})/E({0})$. The model parameters that were used to generate the fake ALMA dataset were again selected by one of the authors and the fake dataset sent to another author who blindly attempted the retrieval assuming that $\epsilon_1=\epsilon_2$. 

We find that, in general, we do not accurately retrieve the ground-truth solution when the underlying $\epsilon_1$ is very different from $\epsilon_2$ and retrievals are performed under the assumption that $\epsilon_1=\epsilon_2$. Because of this, we generate an additional set of forward models corresponding to the ALMA Kalliope dataset presented in this paper, allowing $\epsilon_1$ to be different than $\epsilon_2$. We then use the same procedure outlined in Section \ref{sec:global_TPM} to derive the best-fit values of $\Gamma$, $\epsilon_1$ and $\epsilon_2$. We model the effect of $\epsilon_1$ through the corresponding Fresnel emissivity value between 0 and 1, and allow $\epsilon_2$ to vary between 2 and 60. We find that the retrieved $\epsilon_1$ and $\epsilon_2$ for Kalliope are the same within 2$\sigma$, while the derived value of $\Gamma$ is within the 1$\sigma$ parameter uncertainties whether $\epsilon_1$ and $\epsilon_2$ are coupled or decoupled. However, the solution with $\epsilon_1$ = $\epsilon_2$ ($\Gamma$ = 20$^{+398}_{-8}$ and $\epsilon_1=\epsilon_2$ = 15$^{+5}_{-0}$) has a statistically ($>$ 3$\sigma$) higher $\chi_r^2$ than that for in which $\epsilon_1$ is decoupled from $\epsilon_2$. We therefore adopt the latter as our nominal case, whose results are presented in Section \ref{sec:res}.

\paragraph{Shape model}

We tested the robustness of the solution against the choice of the shape model, which describes the large-scale topography of Kalliope, in two ways. 

As our first shape model test, we performed the TPM retrievals described in Section \ref{sec:global_TPM} using a different shape model, specifically the MPCD-ADAM shape model of \cite{ferrais2022} instead of their SAGE shape model. We find that, when forward models are produced over the same grid of thermal inertia and dielectric constant, the best-fit solution is sensitive to the choice of the shape model but not beyond the level of the uncertainties. The use of the MPCD-ADAM shape model provides retrieved parameter ranges of $\Gamma$ = 156$^{+199}_{-119}$ tiu and $\epsilon$ = 20$\pm$1. This solution overlaps from those we obtain with the SAGE shape model at the 1$\sigma$ level. Importantly, we also find that the minimum $\chi_r^2$ = 16 for the MPCD-ADAM case is more than 3$\sigma$ higher than that obtained with the SAGE shape model. Consistently across thermophysical properties tested, we find that the residual images computed using model images based on the MPCD-ADAM shape model show a dark ring indicating oversubtraction of model flux at the edge of the asteroid due to a shape model that is too large at this orientation. The dark ring is not present when the SAGE model is used. In addition, the MPCD-ADAM shape model is not visually a good match to the ALMA observations. Motivated by this, we adopted the SAGE model as nominal shape model for our analysis.

As our second shape model test, we created a dataset of fake ALMA data images consisting of model images generated using the MPCD-ADAM shape model with added noise as described in Section \ref{sec:test_SNR}, but retrieved the best-fit parameters using the model images generated with the SAGE shape model, following the blind test approach described previously. We find that we can retrieve the ground truth $\epsilon$ at the 3$\sigma$ level in all cases, but not $\Gamma$. Our approach performs better in retrieving the ground-truth $\Gamma$ if the true $\epsilon$ is high ($>$10). This test therefore demonstrates that retrieving an accurate thermal inertia is only possible when a shape model exists that provides a good approximation to the observations. When the shape model does not match the data, the $\chi^2$ value is dominated by regions in the image which contains asteroid in the model but sky in the data, or vice versa. This effect dominates over temperature variations across the surface arising from thermal inertia, and prevents the fits from converging on the true value of thermal inertia.

\paragraph{Electrical skin depth}\label{sec:tests_delec}
We tested the sensitivity of the results to the choice of $\delta_{elec}$. When $\delta_{elec}>>\delta_{th}$, most of the thermal emission arises from the colder subsurface of the asteroid, resulting in a thermal flux that mimics that of a lower emissivity material arising from shallower depths. We explore the robustness of the results to $\delta_{elec}$ by re-running the global model with $\delta_{elec}$= 2 mm at ALMA wavelengths \cite[the value used for M-type asteroid (16) Psyche in][]{dekleer2021,cambioni2022}.

We find that the results of the global model in the case of Kalliope are robust to the choice of $\delta_{elec}$. This finding can be explained by the fact that the electrical skin depth is smaller than the diurnal thermal skin depth in both the nominal and test case. For a low thermal inertia surface, in which the $\delta_{th}$ were smaller, the retrieved parameters may be more sensitive to the choice of $\delta_{elec}$.

\subsubsection{Local model}
\label{sec:local_TPM}

In the local model, we permit the values of $\Gamma$ and $\epsilon$ to vary across the surface of Kalliope, assuming that $\epsilon = \epsilon_1= \epsilon_2$. The local model is described in detail in \citet{cambioni2022} and only briefly summarized here. Differentiating between the effects of $\Gamma$ and $\epsilon$ locally requires resolving temperature as a function of time of day for each surface location. While all datasets presented here cover at least 75\% of Kalliope's rotation period, only the 1.3 mm data have the spatial resolution to map surface properties using this approach.

The local model is based on the concept of a “surface unit”, defined as the subset of facets of the shape model that fall within a given pixel of a single ALMA image. Thus, each ALMA image defines an associated set of surface units, with each unit then occupying different pixels in the ALMA images as the asteroid rotates. We post-process the model images generated for the global model to build thermal emission curves associated with each surface unit, describing how the thermal emission from that unit evolves as a function of time as the asteroid rotates and a facet moves across the plane of sky, and as a function of the local $\Gamma$ and $\epsilon$. We repeat the same procedure to generate the data thermal-emission curves and evaluate the goodness of fit of the model thermal emission curves to the observed thermal emission curves using Eq. \ref{eq:chi_squared} and \ref{eq:chi_2_criterion}. We display the observed and modelled emission, and corresponding surface properties as a function of asteroid-centric longitude and latitude using the Mollweide (equal-area, pseudocylindrical) cartographic map projection with the prime meridian (PM) at the center of the map. To take into account the correlation between nearby pixels that characterizes the ALMA images (each ALMA beam contains about 65 pixels for Kalliope), we bin the surface properties in longitude and latitude on a grid with elements of angular size 5$^\circ\times 5^\circ$ and apply Gouraud shading \citep{gouraud1971continuous}. 

\subsection{Simulating polarized emission}\label{sec:pol}

The fraction of the thermal emission from a surface that is polarized depends on the dielectric constant of the surface and on the emission angle of the radiation, $\theta$ (defined in Section \ref{sec:global_TPM}). If the distribution of emission angles within a telescope resolution element is known, then the polarization fraction can provide a second independent constraint on the dielectric constant. For the case of an asteroid, the surface is modeled as a collection of facets following a choice of shape model, and each facet may have surface roughness that produces a distribution of emission angles at smaller scales. 

We model the expected polarization signal from a smooth surface at the resolution of the observations as follows. We calculate the fractional polarization as a function of emission angle for a given dielectric constant as $F(\theta)=[R_{\perp}(\theta)-R_{\parallel}(\theta)]/[2E(\theta)]$ where the emissivity $E(\theta)=1-[R_{\perp}(\theta)+R_{\parallel}(\theta)]/2$ and $R_{\perp}(\theta)$ and $R_{\parallel}$ are calculated from the Fresnel equations \citep{jackson1975} (see curves in Fig. 5 of \citet{dekleer2021}). We then create simulated images at the same times and pixel scales as the data images, where each pixel is assigned a fractional polarization according to its average emission angle. This fractional polarization image is multiplied by an image corresponding to the best-fit thermal model to produce a simulated polarization flux density image, which is then convolved with the beam. This simulated polarization image is then analyzed in the same way as the actual data to produce curves of polarization fraction as a function of emission angle that account for the shape of the asteroid and the decrease in peak polarization fraction due to averaging in of lower-polarization regions.

 We note that the polarization simulations use scalar fractional polarizations rather than the polarization vectors. This should result in falsely elevated model polarizations near the center of the disk where polarization vectors would otherwise cancel. However, the purpose of the simulation exercise is just to determine the peak polarization signal that should be present in the data for a given dielectric constant, and such signal occurs near the limb where the polarization vectors have the same directionality. Thus this simplification has negligible impact on our conclusions.

\section{Results} \label{sec:res} 

\subsection{Thermophysical and electrical properties of Kalliope}
\label{subsec:TPM_results}

\begin{figure}[ht!]
\centering
\includegraphics[width=16cm]{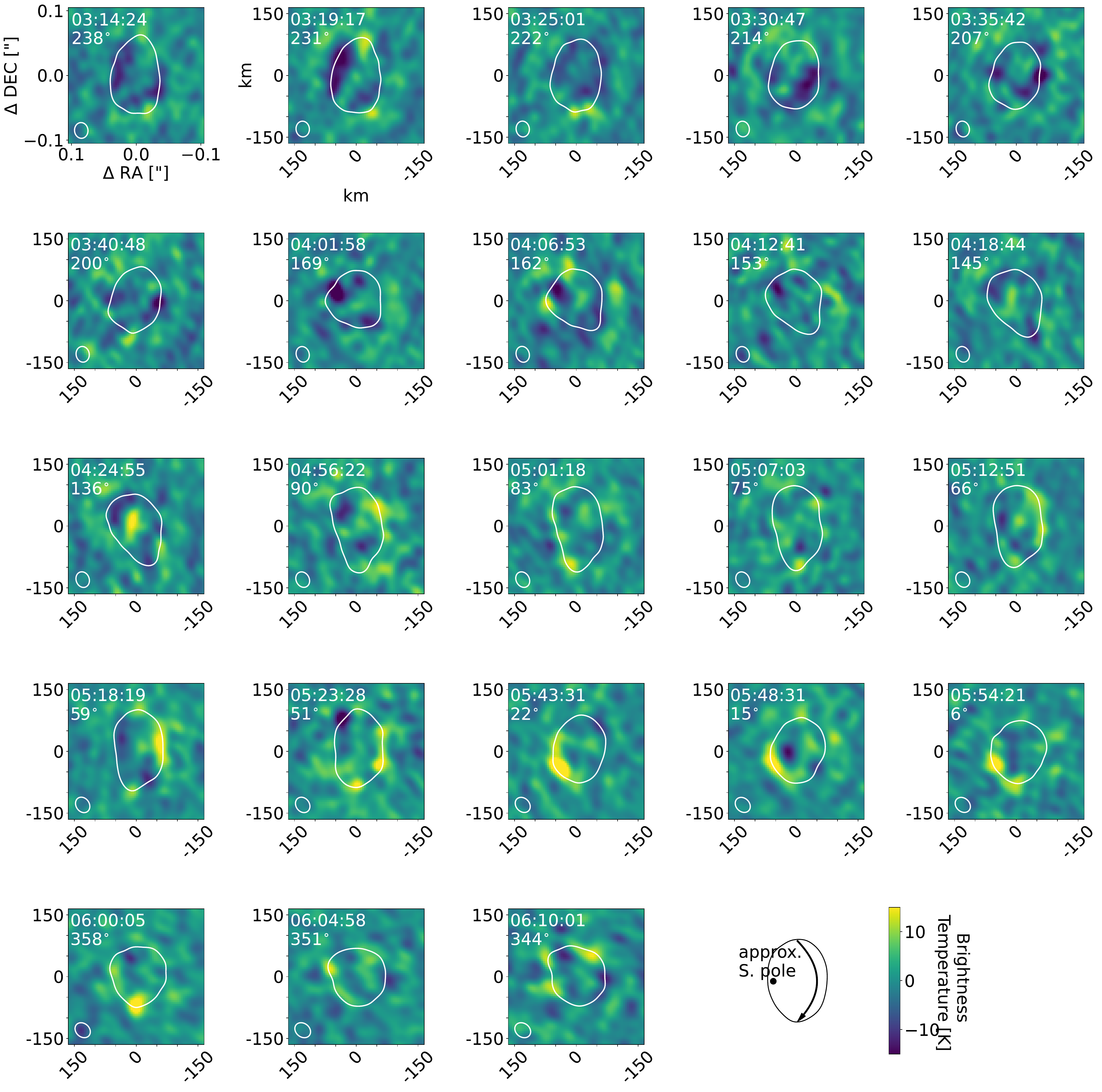}
\caption{Residual images (data minus model) for Kalliope for the ALMA observations. The data image for each frame corresponds to Figure \ref{fig:allims} and the model is the best-fit global model for each epoch. The black outline indicates the shape and location of Kalliope in each frame, and the ellipse in the lower left corner indicates the size and shape of the resolution element. \label{fig:resims_global}}
\end{figure}

Using a model with globally uniform $\Gamma$, $\epsilon_1$ and $\epsilon_2$, we find a best-fit solution of $\Gamma$ = 116$^{+326}_{-91}$, $\epsilon_1$ = 15$^{+2}_{-1}$ and $\epsilon_2$ = 20$^{+0}_{-15}$ for Kalliope at 1.3 mm, corresponding to a minimum $\chi_r^2$ of 12. 
The residual images (data-minus-model) are shown in Figure \ref{fig:resims_global} for the minimum $\chi_r^2$ corresponding to $\Gamma$ = 116, $\epsilon_1$ = 15 and $\epsilon_2$ = 20. Our reported 1$\sigma$ thermal inertia range of 25-440 overlaps with the thermal inertia range of 5-250 that was found for Kalliope based on mid-infrared Spitzer spectroscopy \citep{marchis2012}.

% KdK: Added 11/06/23 -- updated 3/21/24

The 9 and 20 mm data do not have sufficient spatial resolution to provide a strong independent constraint on $\Gamma$ based on the distribution of temperatures across Kalliope's surface. Moreover, the view of Kalliope was nearly pole-on at the time of these observations, which removes the time-of-day temperature variations that motivates the use of a thermophysical model to constrain $\Gamma$. Thus, we compute the thermophysical model for the geometry of the 9 and 20 mm observations using the 1.3-mm best-fit parameter of $\Gamma$ = 116, and compute emissivities at these wavelengths. We find global emissivities for Kalliope, relative to the best-fit thermophysical model, of 0.65$\pm$0.02, 0.56$\pm$0.03, and 0.77$\pm$0.02 at 1.3, 9, and 20 mm respectively. If the emissivity were calculated relative to the model surface temperature without accounting for subsurface emission, the 1.3-mm emissivity would be 0.62$\pm$0.02. In all cases, the electrical skin depth is a small fraction of the effective thermal skin depth, so the surface and subsurface temperatures are comparable; this is still true at 9 and 20 mm only because the pole is facing the sun, which means the dominant period for the thermal wave is not the rotational period but rather the orbital period.

For a smooth, dielectric surface with the low emissivity derived from the 1-9 mm data, the fractional polarization should be tens of \% given the spatial resolution of the 1.3 mm data (based on the computation described in Section \ref{sec:pol}). Instead, no polarization at all is observed, with an upper limit of 1\% at 1.3 mm (Figure \ref{fig:polfrac}) and 2\% at 9 mm (Table \ref{tbl:VLAobs}) after averaging over the full time period of each observation. Achieving a polarization below 1\% in the 1.3 mm data would require a dielectric constant of 1.25 or below, which is at odds with Kalliope's low emissivity. The interpretation of this mismatch is discussed in Section \ref{sec:disc}.

Kalliope's 1.3-mm flux density per unit projected surface area (with areas computed from SAGE model of \cite{ferrais2022}) shows rotational variability at the 10\% level (see Figure \ref{fig:diskint}). The emission per unit area peaks around a sub-observer longitude of 0-30 degrees, and reaches a minimum around 240 degrees (or perhaps less - the viewing geometries corresponding to sub-observer longitudes between 240 and 340 are not covered by the observations).

\begin{figure}[ht!]
\centering
\includegraphics[width=\linewidth]{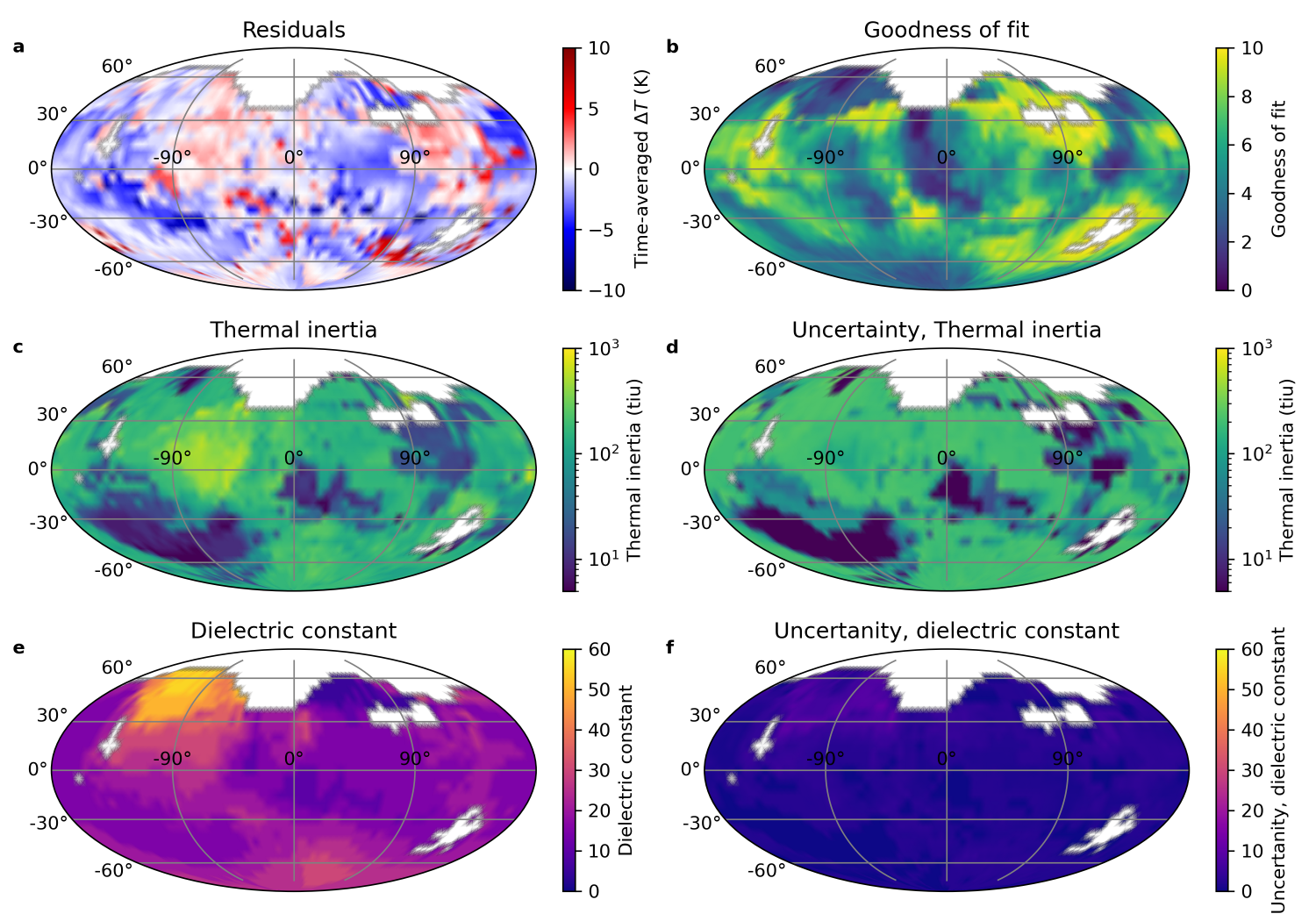}
\caption{Results from the local model. a, time-average of the data minus the model. b, Goodness of fit of the local thermophysical model of Kalliope for the ALMA data. c, Best-fit $\Gamma$. d, Uncertainty in the $\Gamma$. e, best-fit dielectric constant, $\epsilon$. f, uncertainty in dielectric constant, $\epsilon$.}\label{fig:results_local}
\end{figure}

Because the 1.3-mm data are well resolved spatially and temporally, the thermal inertia and dielectric constant can be derived for individual regions on Kalliope's surface, which allows us to explore this rotational variability in more depth. The results for the local model, which fits for $\Gamma$ and $\epsilon$ across the asteroid's surface, are shown in Figure \ref{fig:results_local}. We show the time-average of the residuals (panel a), the goodness of fit in units of $\chi_r^2$ (panel b), the thermal inertia map (best-fit and uncertainty, panels c and d) and the dielectric constant map (best-fit and uncertainty, panels e and f), all projected onto a map of Kalliope's surface in Mollweide projection. The local model has a $\chi_r^2$ between 4 and 6 over most of the surface, thus much lower than that of the global model ($\sim$ 12). The best-fit thermal inertia varies across the surface from $\sim$ 10 to $\sim$ 600–800, although the uncertainties in high-$\Gamma$ regions are large ($\sigma_{\Gamma} \sim$ 100), and no clear statistically-significant trends are seen. The dielectric constant $\epsilon$ is better constrained and its corresponding map (panel e) shows a region with dielectric constant 40–60 in the north between longitudes -150$^{\circ}$ and -120$^{\circ}$ (alternatively, 240 to 270$^{\circ}$). For each surface unit, the results in Figure \ref{fig:results_local} are obtained by computing the median of the model solutions that satisfy Eq. \ref{eq:chi_2_criterion}. This solution is consistent within uncertainty to that obtained by assuming that the combination of $\Gamma$ and $\epsilon$ corresponding to the minimum $\chi^2_r$ in a surface unit is the best-fit solution. We elect to show the local solution based on the median because the resulting thermal inertia map is smoother, so that the appearance of spatial variations that might be artifacts or are not statistically meaningful given the uncertainties is minimized. Both approaches yield the same map of dielectric constant, building confidence in the finding of a region of anomalously high dielectric constant.

\subsection{Kalliope and Linus: A comparison of global properties}

We find rotationally-averaged flux densities for Kalliope of 25.8$\pm$0.9, 0.501$\pm$0.029, and 0.148$\pm$0.006 mJy at 1.3, 9, and 20 mm respectively (Tables \ref{tbl:obs} and \ref{tbl:VLAobs}). For Linus we find flux densities of 1.01$\pm$0.03 mJy at 1.3 mm and 0.021$\pm$0.004 mJy at 9 mm. At 20 mm there is no detection of Linus due to beam dilution by the larger beam; the noise level is a few times the expected brightness of Linus so we cannot place a meaningful upper limit. All uncertainties are 1$\sigma$ and include fit uncertainties as well as flux density scale calibration uncertainties. 

\begin{figure}[ht!]
\centering
\includegraphics[width=16cm]{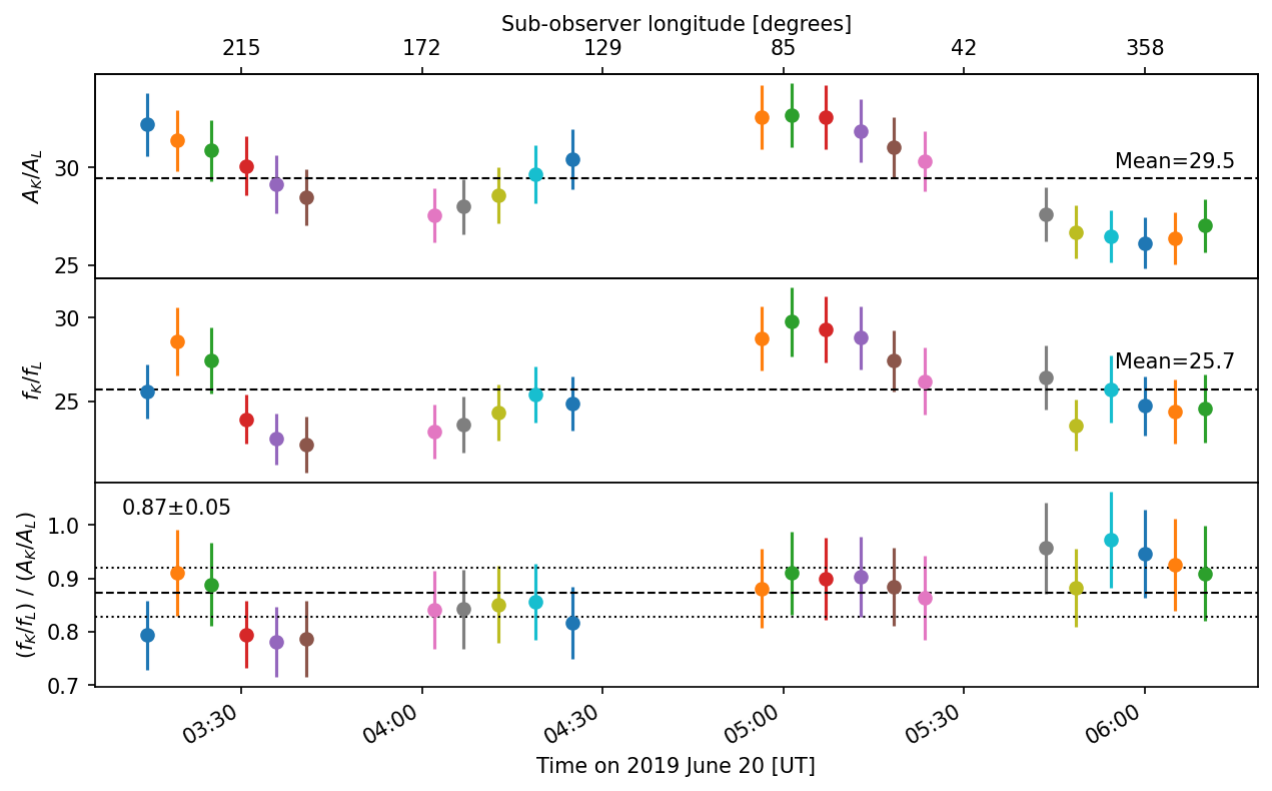}
\caption{Flux density per unit area for Kalliope compared to Linus. (Top) the projected surface area ratio of Kalliope to Linus at the time of each ALMA image; (middle) the flux density ratio of Kalliope to Linus; and (bottom) the ratio of flux density per unit area of Kalliope to Linus. The uncertainties on the area ratio are set by the uncertainties on Linus' diameter alone (and therefore does not represent statistical scatter between points but rather a systematic uncertainty the same across all points). The uncertainties on the flux density are propagated from the measurement uncertainties on both quantities and are statistical scatter. The final stated value of 0.87$\pm$0.05 has an uncertainty that is the quadrature sum of the uncertainties on the mean of the flux density ratio, and the uncertainty on Linus' area. \label{fig:ratios}}
\end{figure}

With ALMA we detect Linus in individual observations; the flux densities of both Kalliope and Linus for each ALMA observation (i.e., jointly imaged set of scans) is listed in Table \ref{tbl:obs}, and the ratio of Kalliope to Linus is plotted in Figure \ref{fig:ratios}. The mean flux density ratio between Kalliope and Linus, $f_K/f_L$, is 25.7$\pm$0.8 for the 1.3 mm observations and 23.9$\pm$4.7 for the 9 mm observations. 

% KdK: Added 08/08/23

These ratios are lower than the ratios of the projected surface areas of the two objects. Linus' size was found to be 28$\pm$2 km based on the drop in system magnitude during Linus eclipse during mutual events \citep{descamps2008}, and was recently refined to 28$\pm$1 km based on a 2021 mutual event campaign \citep{broz2023}. This gives a primary to secondary diameter ratio of 5:1 and area ratio of 29:1 using an effective diameter of 150 km for Kalliope \citep{ferrais2022}. \cite{laver2009} found a relative near-infrared intensity of $\sim$30 for the objects based on direct observations, which they translated into a diameter for Linus of 35$\pm$2 km. However, updating the effective diameter from 166 km (corresponding to the shape model used in their calculation) to 150 km \citep{ferrais2022} gives a diameter for Linus of 28.5 km, consistent with other results and thus also indicating that the near-infrared reflectivity is not substantially different between the objects.

Incorporating the uncertainties on both the sizes and the measurements (and using the instantaneous projected areas from the \cite{ferrais2022} SAGE shape model), we find that the flux density per unit area is lower for Kalliope than Linus by a factor of 0.87$\pm$0.05 at 1.3 mm and 0.66$\pm$0.13 at 9 mm. Either Linus is larger than 28$\pm$1 km \citep{broz2023}, or it has a higher brightness temperature than Kalliope due to a higher emissivity or a higher physical temperature at the depths from which mm-cm emission arises. 

\begin{figure}[ht!]
\centering
\includegraphics[width=16cm]{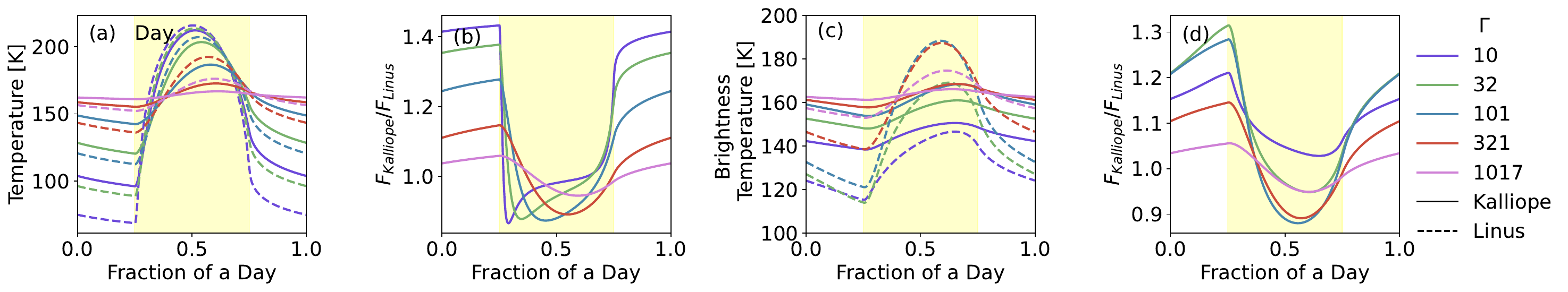}
\caption{Simple thermophysical models for an equatorial point on a spherical asteroid with the rotation rate of Kalliope (4.148 hours) and Linus (3.596 days), demonstrating the change to the diurnal thermal emission curve due to rotation rate alone. (a) Temperature of the surface and (b) ratio of thermal emission from the surface per unit area, and (c) Thermal emission integrated through the subsurface assuming $\delta_{elec}=$10$\lambda$ and converted to brightness temperature, and (d) ratio of thermal emission per unit area including the subsurface emission. Note that because (c) and (d) account for subsurface emission, they also capture the effect of the rotation period on the thermal diurnal skin depth.\label{fig:TPM}}
\end{figure}

We first consider whether Linus' could be warmer than Kalliope due to its longer rotation period alone. For the same thermal inertia, a slower-rotating object will be warmer during the day and cooler at night than a faster-rotating object. Linus should be synchronously rotating based on the short timescale for it to reach such a state \citep{margot2003}. Under that assumption, the rotational period of Linus is 3.596 days, while Kalliope's is 4.148 hours. We explore the quantitative difference to surface temperature that this difference in rotation rate can produce using a simple thermophysical model that gives the temperature of a point on the equator of a spherical asteroid as a function of time of day; the diffusion equation calculation is based on that of \cite{dekleer2021_Ganymede} but is simplified for this application. The resulting temperature curves are shown in Figure \ref{fig:TPM} using thermal inertias from 10 and 1000. We estimate the flux density ratio of Kalliope and Linus due to the difference in rotation alone by taking the ratio of their flux densities at 1.3 mm produced by the modeled surface temperature of (as well as the integrated subsurface emission from) an equatorial surface unit (observed at zero phase angle), integrated across all daytime time steps and weighted by the cosine of the emission angle to account for geometric foreshortening. The minimum ratio of F$_{Kalliope}$/F$_{Linus}$ (per unit surface area) estimated in this way is 0.93, which is achieved for a thermal inertia of $\sim$100. 

% KdK: Added 08/22/23

Thus, while Linus' slower rotation rate does cause a warmer surface, the effect is not strong enough to explain the difference in flux densities between the objects in the 1.3 mm observations. We therefore conclude that the thermophysical or dielectric properties differ between the two targets. A loose anti-correlation has been observed between asteroid size and thermal inertia, which is attributed to the greater ability of large asteroids to generate and retain fine-grained regoliths \citep{delbo2007,hung2022}. Though such a scenario may not apply for two objects that are reaggregated ejecta like Kalliope and Linus may be, this would suggest a higher thermal inertia for Linus and hence a lower daytime temperature, exacerbating the discrepancy with Kalliope rather than resolving it. 

Moreover, the difference between the two objects is even greater at 9 mm, and the observations at that wavelength were made when Kalliope's pole was facing the Sun; assuming alignment of Linus' spin and orbital poles \citep{margot2003}, neither rotation rate nor thermal inertia differences can substantially change the relative brightness temperature between the objects. 

We therefore propose that the difference between the two targets arises from a difference in emissivity. By applying the ratio of flux density per unit area between the targets to Kalliope's measured emissivity, we find that Linus' emissivity is 0.73$\pm$0.04 at 1.3 mm and 0.85$\pm$0.17 at 9 mm.

\section{Discussion} \label{sec:disc}

\subsection{Dielectric and material surface properties of Kalliope}

The radio emissivity of a surface is set by its bulk dielectric constant, which is a function of the dielectric constant of the solid material and the porosity of the bulk surface. The dielectric constant of the solid material is highly sensitive to the metal content, which may be in the form of minerals or of metallic inclusions; the latter scenario requires a much lower metal content to match the observations \citep{dekleer2021}. Based on Kalliope's 1--9 mm emissivities, the upper $\sim$meter of its surface contains a minimum of 45--65 wt.\% metal if the metal is in the form of minerals, and a minimum of $\sim$15--25 wt.\% metal if the metal is in the form of metallic (conducting) inclusions, using the conversions from \cite{dekleer2021}. 

However, the interpretation of the low mm-cm emissivities in terms of the bulk surface dielectric constant is complicated by other apparently conflicting measurements. The amount of emission that should be polarized given this dielectric constant is at the tens of \% level, whereas no polarized emission is detected at all (upper limit of $\sim$1\% after combining all 1.3 mm observations). Kalliope's radar albedo of 0.21$\pm$0.06 \citep[at $\lambda=$12.6 cm and after correcting to its updated size;][]{shepard2015,2007IcarMagri} indicates a metal-poor surface and is apparently incompatible with the 1-9 mm emissivities of 0.56-0.65. However, the higher emissivity at 20 mm of 0.77$\pm$0.02 is compatible within 1$\sigma$ of the radar albedo, suggesting that the low emissivity may in part be the result of a wavelength-dependent phenomenon that suppresses millimeter radiation but not centimeter, discussed below. Interpreting these differences is complicated by the fact that different observations viewed different regions of Kalliope's surface. The ALMA observations viewed sub-observer longitudes of 344--238$^{\circ}$ and sub-observer latitude of -24$^{\circ}$, whereas the VLA data viewed the northern hemisphere nearly pole-on, and the radar data viewed a sub-observer latitude of 26$^{\circ}$N. 

A rough surface can depolarize thermal emission by broadening the range of angles from which the observed emission arises, resulting in a cancellation of polarization vectors. However, roughness alone cannot depolarize the emission to the observed extent \citep{dekleer2021}. The lack of polarization could potentially be explained by the presence of metallic grains $\sim$100 $\mu$m in size that depolarize the emission via scattering, as was proposed for (16) Psyche \citep{dekleer2021}; the similarity between the millimeter emissivities between these two targets is suggestive of a common mechanism. The thermal inertias of the two targets, as derived by resolved millimeter emission, are also similar, with 1$\sigma$ ranges of 80-370 for Psyche and 25-440 for Kalliope. However, the case of (16) Psyche differs from that of Kalliope in that for Psyche, the radar albedo of 0.34$\pm$0.08 \citep{shepard2021} corresponds well with the millimeter emissivity of of 0.61$\pm$0.02 \citep{dekleer2021}. 

Past millimeter and sub-millimeter observations have also found suppressed emissivities for the S-types as well as the M-types, and Vesta notably exhibits these low emissivities at millimeter wavelength but not centimeter, similar to what we see for Kalliope. \cite{keihm2013} show that for several large asteroids including Vesta, their infrared to microwave thermal behavior can be well matched with a low thermal inertia after accounting for subsurface emission, without requiring emissivities below 0.95. However, spatially resolved thermal observations such as those presented here remove much of the degeneracy between thermal inertia and emissivity. At millimeter wavelengths, both M and S type asteroids show a morning to afternoon temperature contrast \citep[e.g., this paper and also][]{dekleer2021,junoalma}. The observed temperature variations are not consistent with the very low thermal inertia that would be required to suppress the emission arising from $\sim$1 cm (i.e. for $\lambda \sim$1 mm observations) sufficiently below the surface temperature to match the low observed emissivity. At 9--20 mm wavelengths, although the data are not well enough resolved spatially to constrain the diurnal temperature variations, the subsurface temperature profile is close to isothermal at the relevant depths due to the pole-on viewing geometry such that subsurface emission cannot account for the low brightness temperatures at these wavelengths either.  

\begin{figure}[t!]
\centering
\includegraphics[width=0.5\linewidth]{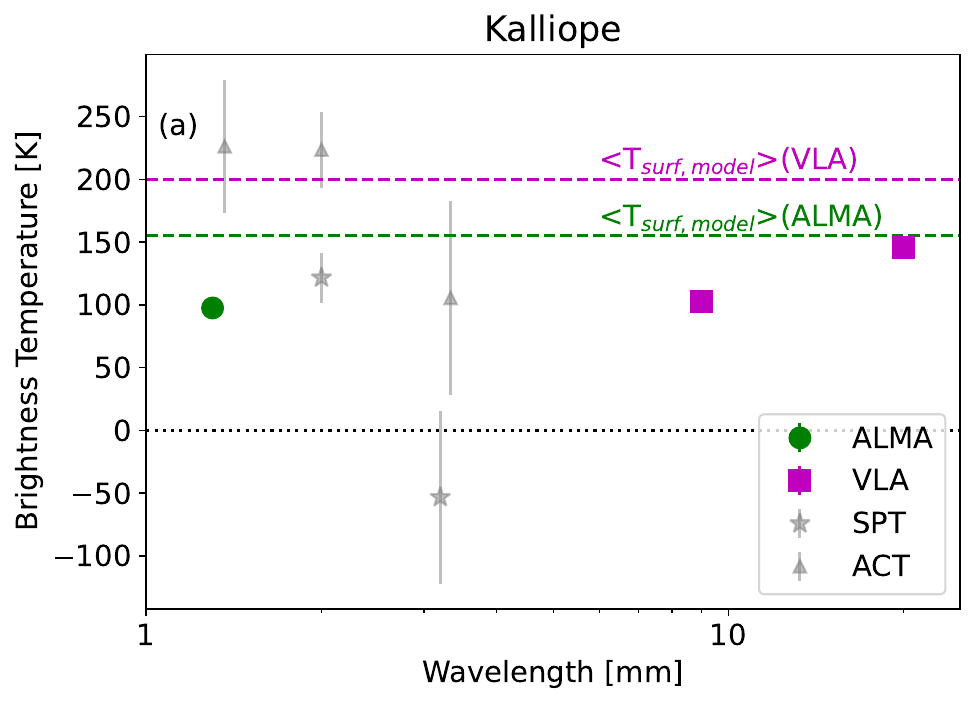}
\caption{Disk-integrated brightness temperature spectrum of Kalliope. All flux densities have been corrected to an asteroid-Sun distance of 2.9112 AU (Kalliope's semimajor axis $a$) using $\sqrt{d/a}$ where $d$ was the separation in AU at the time of the measurement. The conversion to $T_b$ is calculated using the Planck function, and assumes an effective radius of 75 km. The model surface temperatures for the ALMA and VLA data are calculated for $\Gamma=$116, and differ because Kalliope's equator is facing the Sun at the time of the ALMA dataset, while the pole is facing the Sun at the time of the VLA dataset. These lines also therefore represent upper and lower bounds on the surface temperature at other orientations. South Pole Telescope (SPT) data are from \cite{chichura2022} and Atacama Cosmology Telescope (ACT) data from \cite{orlowski2024}, and have been converted from flux density to brightness temperature in the same way as the ALMA and VLA data; these data are aggregated over tens to hundreds of individual measurements that are frequently at the $<1\sigma$ level.}
\label{fig:diskintspecA}
\end{figure}

Since the low emissivity of Kalliope cannot be accounted for by emission from a cold subsurface, it must be due to the material properties of the surface. While models can be used to interpret emissivity in terms of porosity and metal content, simple conversions between these properties cannot account for variations in emissivity with wavelength. Figure \ref{fig:diskintspecA} shows our 1.3, 9 and 20 mm measurements compared to the surface temperature of Kalliope at the time of observation. Kalliope's spectrum is suppressed across mm and cm wavelengths, reaching a minimum at a wavelength of $\sim$millimeters. The trends in emissivity with wavelength may provide important constraints on composition, but lab measurements of the dielectric properties of meteorites and rocks at millimeter wavelengths are lacking. For completeness, we also include measurements from the South Pole Telescope \citep[SPT;][]{chichura2022} and the Atacama Cosmology Telescope \citep[ACT;][]{orlowski2024} on Figure \ref{fig:diskintspecA}. However, we note that these measurements are aggregated over tens to hundreds of individual observations, many of which have a SNR $<$1, and the mismatch between datasets therefore may not be meaningful.

\begin{figure}[t!]
\centering
\includegraphics[width=0.7\linewidth]{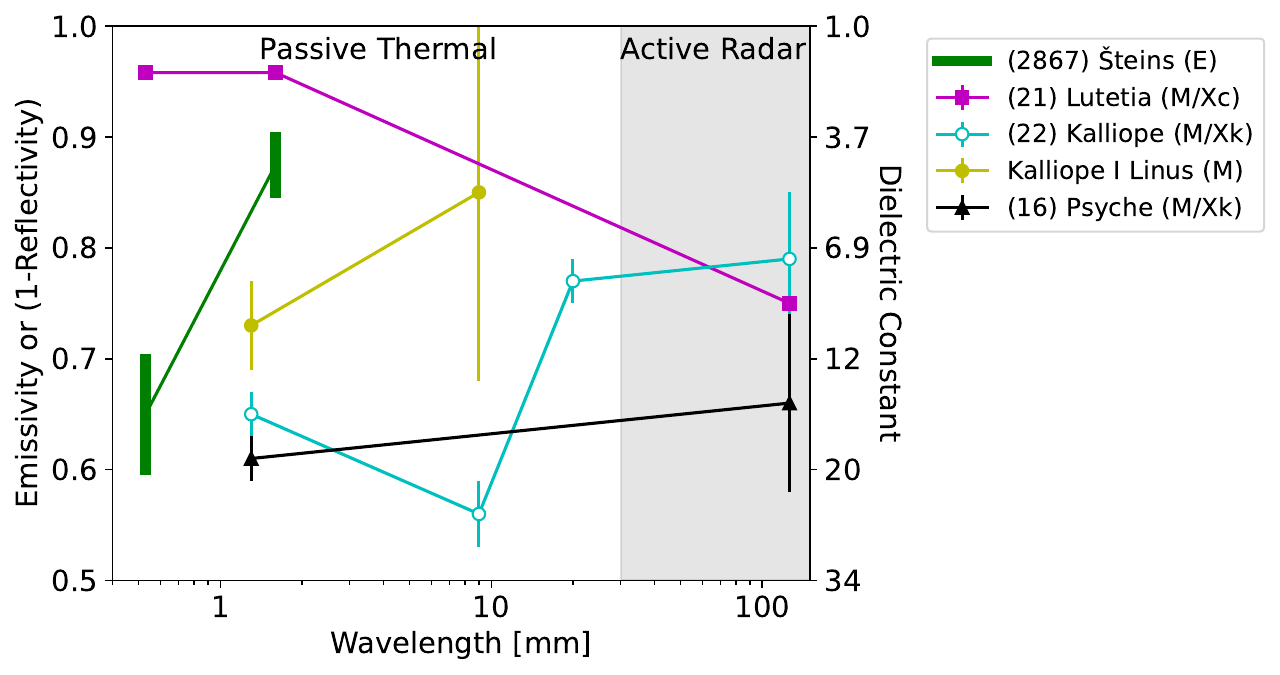}
\caption{Millimeter-centimeter spectra of asteroids that have been observed at high spatial resolution and for which emissivities have been calculated from thermophysical models. (21) Lutetia and (2867) \v{S}teins millimeter data are from the MIRO instrument on the \textit{Rosetta} spacecraft \citep{gulkis2010,gulkis2012}; 1.3 mm data on (16) Psyche are from \cite{dekleer2021}; radar data are from \cite{2007IcarMagri,shepard2015,shepard2021} and Kalliope's radar albedo has been corrected to its updated diameter. Asteroids are designated by spectral type in the Tholen and Bus-DeMeo classification schemes \citep{tholen1984,bus1999,demeo2009}.}
\label{fig:diskintspecB}
\end{figure}

In the absence of lab measurements, we can compare to other asteroids for which robust multi-frequency measurements exist. While mm-cm spectra of asteroids have been previously presented and provide valuable information \citep{redman1992,redman1998}, high spatial resolution observations provide simultaneous constraints on shape, size, and thermal inertia, removing several sources of degeneracy in interpretation and providing a more robust spectrum of the emissivity of the surface materials. Among other targets for which the emissivity of surface materials has been measured in this way, the Rosetta/MIRO measurements of (21) Lutetia and (2867) \v{S}teins provide a comparison point, shown on Figure \ref{fig:diskintspecB}. Both (21) Lutetia and Kalliope are classified as M types (alternatively, Xc and Xk respectively), and have radar albedos that are the same to within uncertainties. (21) Lutetia was found to have a very low thermal inertia of $<$20, increasing with depth after the upper few cm, and an emissivity of $>$0.9 \citep{gulkis2012}. Measurements of (4) Vesta, (1) Ceres, (2) Pallas, and (10) Hygiea could also be matched by an emissivity close to 1 across a broad wavelength range after interpretation with a thermophysical and radiative transfer model analogous to what we employ here \citep{keihm2013}. In contrast, the $\sim$5-km E-type asteroid (2867) \v{S}teins was found to have emissivities of 0.6-0.7 and 0.85-0.9 at 0.53 and 1.6 mm, respectively, corresponding to dielectric constants in the range of 4-20 \citep{gulkis2010}. The low emissivity at 0.53 mm and steep spectral slope remain unexplained. (2867) \v{S}teins' millimeter thermal inertia of 450-850 measured by MIRO is also substantially higher than its thermal inertia of 150$\pm$60 measured in the mid-infrared by Spitzer \citep{lamy2008}. 

It is clear that the IR -- cm thermal spectra of asteroids show distinctive trends and therefore have the potential to provide constraints on surface properties and composition across asteroid spectral classes. However, deriving true material emissivity spectra requires accurate shape and rotation models, an accurate thermal inertia, and interpretation with a thermophysical and radiative transfer model. The set of asteroids that these conditions apply to remains small. Future lab measurements of the dielectric properties of meteorites at sub-mm through centimeter wavelengths, combined with observed mm-cm emissivity spectra of asteroids across a range of spectral types, have the potential to provide new a new tool for assessing the composition and properties of asteroid surfaces across classes.

\subsection{Heterogeneities in Kalliope's surface properties}

We do not find strong evidence for heterogeneity in thermal inertia across Kalliope's surface, given the large uncertainties. The strongest evidence for heterogeneity in Kalliope's material properties is the region of high dielectric constant (low emissivity) around longitudes 210--280$^{\circ}$ in the north. This may be a region that is less porous and/or higher in metal content. The emissivity in this region is $\sim$0.45, which is only achievable assuming the metal is present in minerals if the metal content is 90 wt.\% or higher and the porosity is under 10\%, an unlikely scenario. Under the conducting inclusion model, this low emissivity can be achieved with a metal content of at least 35 wt.\% and a porosity of $\lesssim$30\%. 

This region of high dielectric constant is only observed when it is at Kalliope's limb due to our incomplete rotational coverage. A shape model mismatch in this region could therefore produce a similar effect. However, the robustness of the low emissivity in this region is supported by the rotationally-resolved radar albedos, which are maximal at this location. Radar observations were made with a sub-Earth latitude of 26$^{\circ}$N and sub-Earth longitudes of 30, 70, 269, and 350$^{\circ}$ \citep[][and C. Magri, personal communication]{2007IcarMagri}. The observation at 269$^{\circ}$ longitude is therefore centered on our low-emissivity region, and finds the highest radar albedo of the four (0.23). Conversely, the radar observations centered on 30 and 70$^{\circ}$, which align with the region of Kalliope where we measure the highest emissivity, measured the lowest radar albedos of 0.12 and 0.17. The radar observation centered on 350$^{\circ}$ covers a hemisphere with average millimeter emissivity, and correspondingly has an intermediate radar albedo of 0.20. The observed radar variability is statistically significant given the 1$\sigma$ uncertainties of 0.01 in radar albedo (note that this does not include an overall flux calibration uncertainty of 25\%). The anti-correlation between radar albedo and millimeter emissivity across Kalliope's surface supports the presence of true spatial variations in the dielectric constant and, by extension, metal content.

\subsection{History of Kalliope and Formation of Linus}

Linus' orbit suggests that it formed as reaggregated material following an impact on Kalliope, and has been gradually moving outwards from a formation location close to Kalliope due to tidal dissipation \citep{margot2003}. Its circular orbit implies that tidal dissipation is more effective in Linus than in Kalliope. 

At the time of Linus' discovery, Kalliope's density was found to be 2.37$\pm$0.4 g cm$^{-3}$, too low for a metallic object. However, Kalliope's diameter has since been revised from the 181 km used to calculate the initial density to 150 km today \citep{ferrais2022}; the new density, combined with a refined orbit for Linus based on a longer interval of observation, results in a bulk density of 4.40$\pm$0.46 g cm$^{-3}$, within uncertainties of the bulk density of 3.78$\pm$0.34 g cm$^{-3}$ for (16) Psyche averaged across several studies \citep{2022SSRvElkinsTanton}.

This high bulk density, combined with the surface density of 2.3 g cm$^{-3}$ inferred from its radar albedo \citep{shepard2015}, led \cite{broz2022} to conclude that Kalliope is a differentiated body, most of whose mantle was ejected following the impact that produced Linus. Hydrodynamic simulations of the impact process show that Kalliope could be core-dominated, with a large iron core constituting most of the volume and off-center within the body, and raised areas on Kalliope's shape attributed to reaccumulation of ejecta fragments \citep{broz2022}.

However, the surface density derived from the millimeter emissivity is 4.2 g cm$^{-3}$ \citep[using the conversion from][for direct comparison with radar-derived densities]{ostro1985}, matching the bulk density. If the radar-derived densities were excluded, the millimeter and bulk densities together would imply a uniform composition and porosity, appropriate for an undifferentiated body or else a reaggregated mixture of fragments from a differentiated parent body. A determination of whether Kalliope's surface density is indeed much lower than its bulk density therefore requires understanding the reason for the strong wavelength-dependence in its emissivity and reflectivity at mm-cm wavelengths.

Further constraints on Kalliope's history come from infrared spectroscopy. Although featureless in the near-infrared, Kalliope exhibits a shallow ($\sim$3\% band depth) feature at 0.9 $\mu$m, consistent with the presence of low-Fe pyroxenes \citep{2004AJClark,2010IcarFornasier}, as well as 3 $\mu$m and 0.43 $\mu$m attributed to aqueous alteration products \citep{2010IcarFornasier,rivkin2000}. Hypotheses for the composition of the M-types that exhibit the 0.9 $\mu$m feature include core material with a veneer of low-Fe pyroxene, either from a mantle that was mostly lost by impacts or alternatively from silicate infall; products of a carbon - olivine reaction producing enstatite, metallic iron, and silica \citep{hardersen2005}; or analogs to CB carbonaceous chondrites, which are breccia containing extensive metallic inclusions. Additional data that now need to be matched by an origin scenario are (a) the fact that the newly-identified family members are classified as M-types \citep{broz2022}; (b) the low emissivities and depolarization we observe in the millimeter, which suggest metal content in the form of metallic inclusions in a chemically-reduced regolith; (c) the apparently lower metal content in Linus compared to Kalliope, despite the fact that both are featureless in the near-infrared \citep{laver2009}; and (d) the evidence that there is a localized region on Kalliope with higher metal content than the average surface.

The apparently lower metal content of Linus compared to Kalliope, and the heterogeneous distribution of metal across Kalliope itself, are consistent with the scenario in which Kalliope is a differentiated body which contains a higher ratio of metal/silicate compared to Linus. However, \cite{cloutis1990} show that in a metal-silicate mixture, olivine is detectable at 1 $\mu$m if it is present at $>$25\%, while ferrous pyroxene is detectable if it is present at $>$10\%. Thus the fact that neither Kalliope nor Linus exhibits spectral features in the near-infrared, and that all confirmed family members are also M-types, means that the mantle material cannot have a significant component of olivine or ferrous pyroxene. Mg-silicates such as enstatite, in contrast, are spectrally featureless, and changing the metal/enstatite ratio would change the millimeter emissivity while maintaining a featureless visible-near-infrared spectrum. Collectively, the observations suggest a scenario where Kalliope is the remnant of a differentiated body that formed under reducing conditions such that the silicates are mainly present in the form of Mg minerals, leaving the Fe in pure metallic form. In that scenario, $\sim$15-25 wt.\% metal is sufficient to explain the observations, an amount fully consistent with enstatite chondrites or even some carbonaceous chondrites without requiring enhanced metal content. Kalliope was disrupted at least once and reaggregated with core material distributed heterogeneously across its surface, while Linus is an aggregation of collisional fragments that has a lower component of core material than Kalliope itself.
\section{Conclusions} \label{sec:conc}
We observed Kalliope and its moon Linus at 1.3, 9, and 20 mm wavelengths with ALMA and the VLA over the majority of Kalliope's rotation period. The 1.3 mm data resolve Kalliope at $\sim$30 km resolution, and both the 1.3 and 9 mm data detect thermal emission from Linus in addition to Kalliope. We use the temperature variations across the surface of Kalliope over its rotation to fit a thermophysical and radiative transfer model, finding a thermal inertia of 116$^{+326}_{-91}$ and a dielectric constant of 15$^{+2}_{-1}$.

Kalliope's emissivity is 0.65$\pm$0.02, 0.56$\pm$0.03, and 0.77$\pm$0.02 at 1.3, 9, and 20 mm respectively; it is suppressed at mm wavelengths compared to cm. Linus' surface exhibits a higher emissivity than Kalliope's, with millimeter emissivities of 0.73$\pm$0.04 and 0.85$\pm$0.17 at 1.3 and 9 mm respectively, indicative of a lower metal content and/or a higher surface porosity than Kalliope. Kalliope also exhibits a localized region, roughly between longitudes 210 and 280$^{\circ}$ in the northern hemisphere, with an emissivity of $\sim$0.45, indicative of a heterogeneous distribution of metal across its surface. This interpretation is supported by the higher radar albedo at these longitudes.

Surface densities derived from the millimeter observations ($\sim$4 g/cm$^3$) disagree with cm- and radar-derived densities ($\sim$2-3 g/cm$^3$), and hence it is unclear whether Kalliope's surface density is truly lower than its bulk density. Nevertheless, the heterogeneous distribution of metal across Kalliope's surface, combined with the lower metal content of Linus compared with Kalliope, support the interpretation that Kalliope is a disrupted differentiated object of which Linus is a fragment that has a higher mantle/core ratio than Kalliope itself. However, the evidence for only a minimal presence of Fe-bearing silicates on Kalliope, Linus, or any family member indicates a very Fe-poor mantle, suggesting that Kalliope formed under reducing conditions. This is supported by the depolarization of the mm thermal emission we observe, which suggests metallic particles in the near-surface regolith in the size range of tens of $\mu$m to a few mm. In this case, the millimeter emissivities can be matched with a metal content of only $\sim$15-25 wt.\%. An Fe-poor silicate regolith with $\sim$tens of \% metal grains in consistent with meteorite classes such as enstatite chondrites or some carbonaceous chondrites without requiring an enhancement in metal content.

These observations demonstrate that mm--cm data can distinguish variations in surface properties and composition between objects that are spectrally indistinguishable in the visible and near-infrared, providing a unique avenue for constraining the compositions of reduced and metal-rich bodies that lack spectral features at other wavelengths. High spatial resolution thermal data also directly resolve asteroid shape and time-of-day temperature variations, removing degeneracies that are present when interpreting unresolved thermal observations. The distinctive emissivity spectra of asteroids across the mm--cm, with clear emissivity minima occurring at different wavelengths for different targets, suggests that such spectra contain another potential source of information. Future laboratory measurements of analog materials across sub-mm to cm wavelengths, combined with spatially-resolved mm--cm observations of asteroids across a range of spectral classes, would allow us to better calibrate mm--cm thermal spectra as a tool for studying the surfaces of these objects.

\begin{acknowledgments}
\textbf{Acknowledgements:} The authors thank Chris Magri for providing the radar cross-sections from individual dates of observation in 2001. KdK is grateful to the staff at the North America ALMA Science Center (NAASC) for their support through a PI data reduction visit. The authors are grateful to J. Orlowski-Scherer for providing the ACT flux density measurements and assistance in their interpretation. KdK acknowledges support from The Research Corporation through Cottrell Scholar Award \#2024-119. KdK and SC were supported by grant \#2019-1611 from the Heising Simons Foundation.  SC acknowledges the Crosby Distinguished Postdoctoral Fellowship of Department of Earth, Atmospheric and Planetary Sciences, Massachusetts Institute of Technology (MIT). This paper makes use of the following ALMA data: ADS/JAO.ALMA\#2018.1.01271.S. ALMA is a partnership of ESO (representing its member states), NSF (USA) and NINS (Japan), together with NRC (Canada), NSTC and ASIAA (Taiwan), and KASI (Republic of Korea), in cooperation with the Republic of Chile. The Joint ALMA Observatory is operated by ESO, AUI/NRAO and NAOJ. This paper makes use of VLA data from program \#20B-259. The National Radio Astronomy Observatory is a facility of the National Science Foundation operated under cooperative agreement by Associated Universities, Inc. The Arecibo Observatory was part of the National Astronomy and Ionosphere Center, which was operated by Cornell University under a cooperative agreement with the National Science Foundation. Part of this research was conducted at the Jet Propulsion Laboratory, California Institute of Technology, under contract with the National Aeronautics and Space Administration (NASA). This material is based in part upon work supported by NASA under the Science Mission Directorate Research and Analysis Programs.
\end{acknowledgments}

\clearpage

\bibliography{kalliope}{}
\bibliographystyle{aasjournal}

\end{document}